\documentclass[twocolumn]{aastex701}
\usepackage{amsmath}
\usepackage{subcaption}

\newcommand{\I}{\ensuremath{\,\textrm{F814W}}}
\newcommand{\V}{\ensuremath{\,\textrm{F606W}}}
\newcommand{\T}{\ensuremath{\,\textrm{TRGB}}}
\newcommand*{\degree}{\ensuremath{^\circ}}
\newcommand*{\CVnC}{CVn\,C}

\begin{document}

\title{A Quenched and Relatively Isolated Dwarf Galaxy in the Local Volume}

\author[0009-0005-9382-1362]{Tehreem N. Hai}
\email{tnh57@physics.rutgers.edu}
\altaffiliation{NSF Graduate Research Fellow}
\affiliation{Department of Physics and Astronomy, Rutgers University, Piscataway, NJ 08854, USA}

\author[0000-0001-5538-2614]{Kristen B. W. McQuinn}
\email{kmcquinn@stsci.edu}
\affiliation{Department of Physics and Astronomy, Rutgers University, Piscataway, NJ 08854, USA}
\affiliation{Space Telescope Science Institute, 3700 San Martin Drive, Baltimore, MD 21218, USA}

\author[0000-0002-1200-0820]{Yao-Yuan Mao}
\email{yymao@astro.utah.edu}
\affiliation{Department of Physics and Astronomy, University of Utah, Salt Lake City, UT 84112, USA}

\author[0000-0002-2970-7435]{Roger E. Cohen}
\email{rc1273@physics.rutgers.edu}
\affiliation{Department of Physics and Astronomy, Rutgers University, Piscataway, NJ 08854, USA}

\author[0000-0003-3408-3871]{David Shih}
\email{shih@physics.rutgers.edu}
\affiliation{Department of Physics and Astronomy, Rutgers University, Piscataway, NJ 08854, USA}

\author[0000-0002-9599-310X]{Erik Tollerud}
\email{etollerud@stsci.edu}
\affiliation{Space Telescope Science Institute, 3700 San Martin Drive, Baltimore, MD 21218, USA}

\author[0009-0005-8590-0829]{Joseph A. Breneman}
\email{jab964@physics.rutgers.edu}
\affiliation{Department of Physics and Astronomy, Rutgers University, Piscataway, NJ 08854, USA}

\author[0000-0001-8416-4093]{Andrew E. Dolphin}
\email{adolphin@rtx.com}
\affiliation{Raytheon, 1151 E. Hermans Road, Tucson, AZ 85756, USA}
\affiliation{Steward Observatory, University of Arizona, 933 N. Cherry Avenue, Tucson, AZ 85719, USA}

\author[0000-0002-8092-2077]{Max J. B. Newman}
\affiliation{Space Telescope Science Institute, 3700 San Martin Drive, Baltimore, MD 21218, USA}
\email{mjbnewman25astro@gmail.com}

\author[0000-0003-2599-7524]{Adam Smercina}\thanks{NHFP Hubble Fellow}
\email{asmercina@stsci.edu}
\affiliation{Space Telescope Science Institute, 3700 San Martin Drive, Baltimore, MD 21218, USA}

\begin{abstract}

An increasing number of discoveries of isolated and quenched dwarf galaxies are challenging the idea that the present-day local environment of low-mass systems is the main determinant of their quenching. We present new Hubble Space Telescope (HST) data of one such system, the dwarf galaxy Canes Venatici C (\CVnC{}). \CVnC{} is a low-mass ($3.4^{+4.2}_{-2.6}\times10^6\: M_{\odot}$) galaxy with a Tip of the Red Giant Branch distance of $8.43^{+0.47}_{-0.32}$ Mpc determined from the resolved stars in the HST imaging, which we also use to derive \CVnC{}'s structural parameters. \CVnC{}'s distance places \CVnC{} in the Local Volume and in an isolated environment with the most tidally influential $L_{\star}$ galaxy $\rm>5R_\text{vir}$ away. Additional constraints from the HST color-magnitude diagram, archival Far-Ultraviolet (FUV), and neutral hydrogen (H\textsc{i}) data show that \CVnC{} is quenched, with no evidence of star formation in the last 100 Myr and no detectable gas ($\rm M_{\text{H}\textsc{i}}<1.5\times10^6\:M_{\odot}$). Circumstantial evidence suggests that \CVnC{} may have quenched via past interactions with the $L_{\star}$ galaxy NGC 4631 ($L_{K}=10^{10.4}\:L_{\odot}$), and was possibly sent on an extreme backsplash orbit by the tidal dissolution of a subhalo group. However, other quenching mechanisms---such as stripping via the cosmic web---cannot be ruled out.
\CVnC{} adds to the growing number of quenched dwarf galaxies in under-dense environments, a population that will be critical to defining the mass and environment regimes in which different quenching mechanisms operate.

\end{abstract}


\keywords{Galaxy environments (2029), Stellar populations (1622), Dwarf galaxies (416), Galaxy quenching (2040), HST photometry (756), Hertzsprung Russell diagram (725)}

\section{Introduction} \label{sec:intro}

The present-day local environment of dwarf galaxies has long been considered a dominant driver of their star formation (or lack thereof). This is supported by the observation of the morphology-density relation, in which dwarf galaxies near a massive host galaxy (i.e., satellites) tend to have quenched star formation and low gas fractions, whereas low-mass systems outside the influence of a massive host (i.e., field dwarf galaxies) tend to be star-forming and gas-rich \citep[e.g.,][]{Geha2012,Spekkens2014,Putman2021}. This morphology-density relation can be physically explained by the environmental effects experienced by satellites around a massive galaxy; ram-pressure stripping while passing through a massive host's halo can strip a significant amount of the satellite's gas and quench it, while tidal stripping of the dwarf galaxy at close pericentric distances to the host can also remove gas, as well as the dark matter and stars, from the system \citep{Simpson2018}. In contrast, galaxies which presumably evolve without such environmental influences (i.e., field galaxies) are more likely to retain their gas and remain star-forming until the present day.

However, a number of recent discoveries of isolated dwarf galaxies with no discernable recent star formation suggests that factors other than the density of the present-day local environment may play a role in whether a low-mass galaxy ceases forming stars \citep[e.g.,][]{Roman2019,Polzin2021,Prole2021,Sand2022,Casey2023,Carleton2024,Li2024,Bidaran2025,Jones2025}\footnote{We note that the cited studies use a variety of definitions of quenching and quiescence in star formation, including evidence from galaxy colors, lack of H$\alpha$ or ultraviolet emission, and star formation histories inferred from spectral fitting.}. 

It is still an open question what the quenching mechanisms of these isolated dwarf $( M_*\lesssim10^9\:M_{\odot})$ galaxies might be; some theories point to the environment of the galaxy, but go beyond the present-day host halo influence. 
For example, the environment around a dwarf galaxy at earlier times may be critical in quenching. This is supported by observations of a few dwarf galaxies which show signatures of ram-pressure and tidal stripping even when found outside the virial radius ($R_\text{vir}$) of a massive host galaxy; a number of these systems may be `associated' galaxies, which are thought to have entered and left the host halo in the past \citep[e.g.,][]{Sales2007,Teyssier2012,Santos-Santos2023,Bhattacharyya2024,Wei2025}. While passing through the host halo, these galaxies can lose gas and quench via ram-pressure and tidal stripping, similar to the environmental processing experienced by conventional satellite galaxies \citep{Simpson2018}. Another possible environmental quenching mechanism is pre-processing (e.g., via ram-pressure stripping in low-mass satellite group environments); this can strip the gas from dwarf galaxies even before infall into the virial radius of a massive host \citep{Samuel2023}. Finally, the large-scale environment around dwarf galaxies may also be important; ram-pressure from the cosmic web can both strip gas from these systems \citep{Benitez-Llambay2013} and reignite their star formation \citep{Wright2019, Liao2019}.

Internal processes may also play a role in quenching low-mass galaxies. Stellar feedback-driven outflows can expel neutral gas and aid quenching in low-mass galaxies \citep{El-Badry2016,Rey2020,Jones2025}, while feedback from nearby galaxies may also facilitate star formation by compressing the dwarf galaxy's gas \citep{Wright2019}. 

In addition to environment and internal processes, global processes also influence star formation at the lowest galaxy mass scales ($\rm M_* <10^5 \: M_{\odot}$); reionization may have heated the gas in these ultra-faint dwarf galaxies and prevented them from accreting further gas, quenching their star formation \citep{brown2014,Wheeler2015,Rey2020}.

In light of myriad mechanisms which impact the ability of dwarf galaxies to retain gas and continue forming stars, a large sample of galaxies in different environments with precisely measured properties is needed to disentangle the relative effects of environmental and internal processes. Here we add one such galaxy to this emerging sample: Canes Venatici C (\CVnC{}). Using high-resolution Hubble Space Telescope (HST) imaging, we determine that \CVnC{} is a dwarf galaxy ($\rm M_*\sim3\times 10^6 \:M_{\odot}$) located at a distance of $D=8.4$ Mpc, at a separation of more than 5$R_\text{vir}$ from its most tidally influential $L_{\star}$ galaxy neighbor. \CVnC{} has no detectable gas ($\rm M_{\text{H}\textsc{i}}<1.5\times10^6\:M_{\odot}$) and has experienced little to no star formation in the last 1 Gyr, indicating that it is a quenched dwarf galaxy in an isolated environment.

The organization of this paper is as follows. Section \ref{sec:obs_data} describes our data and the methods used to extract photometry of \CVnC{}'s resolved stellar population. Section \ref{sec:results} details the results of our analysis, including structural parameters (Section \ref{sec:struc_params}),  the color-magnitude diagram (CMD) of stars in \CVnC{} (Section \ref{sec:cmd}), our measurement of the distance to \CVnC{} using the Tip of the Red Giant Branch (TRGB) method (Section \ref{sec:trgb}), constraints on \CVnC{}'s recent star formation and gas content (Section 
\ref{sec:sfh}), our measurement of \CVnC{}'s integrated luminosity and present-day stellar mass (Section \ref{sec:lum_mass}, and the environment around \CVnC{} (Section \ref{sec:envir}). Section \ref{sec:discuss} discusses plausible quenching mechanisms for \CVnC{}, and Section \ref{sec:conclusion} summarizes our results and possible future directions.

\section{Observations and Data Reduction} \label{sec:obs_data}

\CVnC{} was detected in DESI Legacy Imaging Survey data, similar to the nearby ultra-faint dwarf galaxies Pegasus~W, Leo~M, and Leo~K \citep{McQuinn2023,McQuinn2024LeoMK}. \CVnC{} was named after the constellation it resides in, Canes Venatici, plus a letter---as has been done with other nearby galaxies (e.g., Leo~T, Leo~P, Pegasus~W, Leo~M, Leo~K). Other identifiers for \CVnC{} include BTS~113 \citep{Binggeli1990}, KKH~78 \citep{Karachentsev2001}, and SMDG J1217443+332043 \citep{Karunakaran2024}.

\subsection{Observations} \label{sec:obs}

We obtained follow-up optical imaging of \CVnC{} with the HST Advanced Camera for Surveys Wide Field Camera (ACS/WFC; \citealp{Ford1998}), and parallel imaging with the HST Wide Field Camera 3 (WFC3) UVIS, as part of Cycle 31 observations under HST-GO-17481 (PI McQuinn). All HST data associated with \CVnC{} is available in MAST: \dataset[10.17909/vvhf-ad76]{https://doi.org/10.17909/vvhf-ad76}.

The primary data for \CVnC{} consisted of ACS WFC observations of the stellar population in the F606W and F814W filters,  which enable the construction of a high-fidelity CMD and facilitate measurements of the distance and star formation history (SFH) of the system. The ACS/WFC observations were conducted on January 12th, 2024; two observations were taken per filter in one orbit, with a total exposure time of 1020 s per filter. To aid in the rejection of cosmic rays and remove detector defects, we used a $5\times5$ pixel dither (\texttt{acs wfc dither line pattern \#14}) between the two observations. The ACS field of view was centered on J2000 RA = 12:17:39.437, Dec = +33:20:49.68. This was slightly off-center from the galaxy, and was chosen to maximize the coverage of \CVnC{} within the ACS FOV while avoiding bright stars.

In order to quantify the level of background and foreground contamination expected in the primary data, we used WFC3 with the F606W and F814W filters for parallel observations.
The parallel field was not ultimately used in our analysis because the galaxy's small angular extent fit within $\sim$1 of the ACS chips, allowing us to use off-target regions in the ACS data as representative of potential background contamination.

\begin{figure*}[t!]
    \begin{subfigure}[t]{0.5\textwidth}
        \centering
        \includegraphics[scale=0.4,alt={Two panels showing HST images (Declination vs.\ Right Ascension) of the ACS and WFC3 observations of \CVnC{}. The stellar disk of the galaxy fits within one of the two ACS chips and shows a smooth, diffuse light distribution in the ACS image. Both the ACS and WFC3 fields show minimal foreground and background contamination.}]{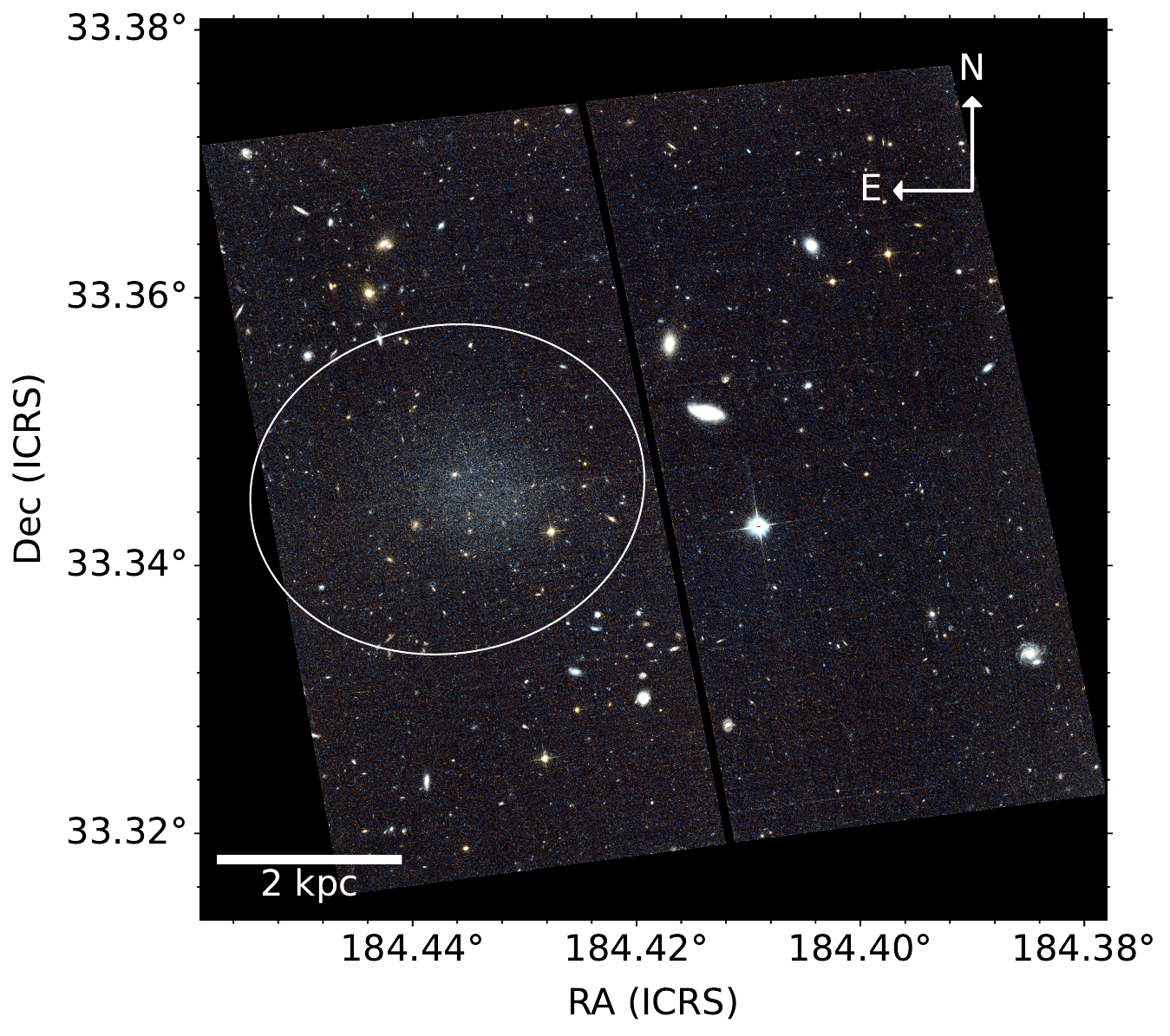}
    \end{subfigure}%
    ~
    \begin{subfigure}[t]{0.2\textwidth}
        \centering
        \includegraphics[scale=0.4]{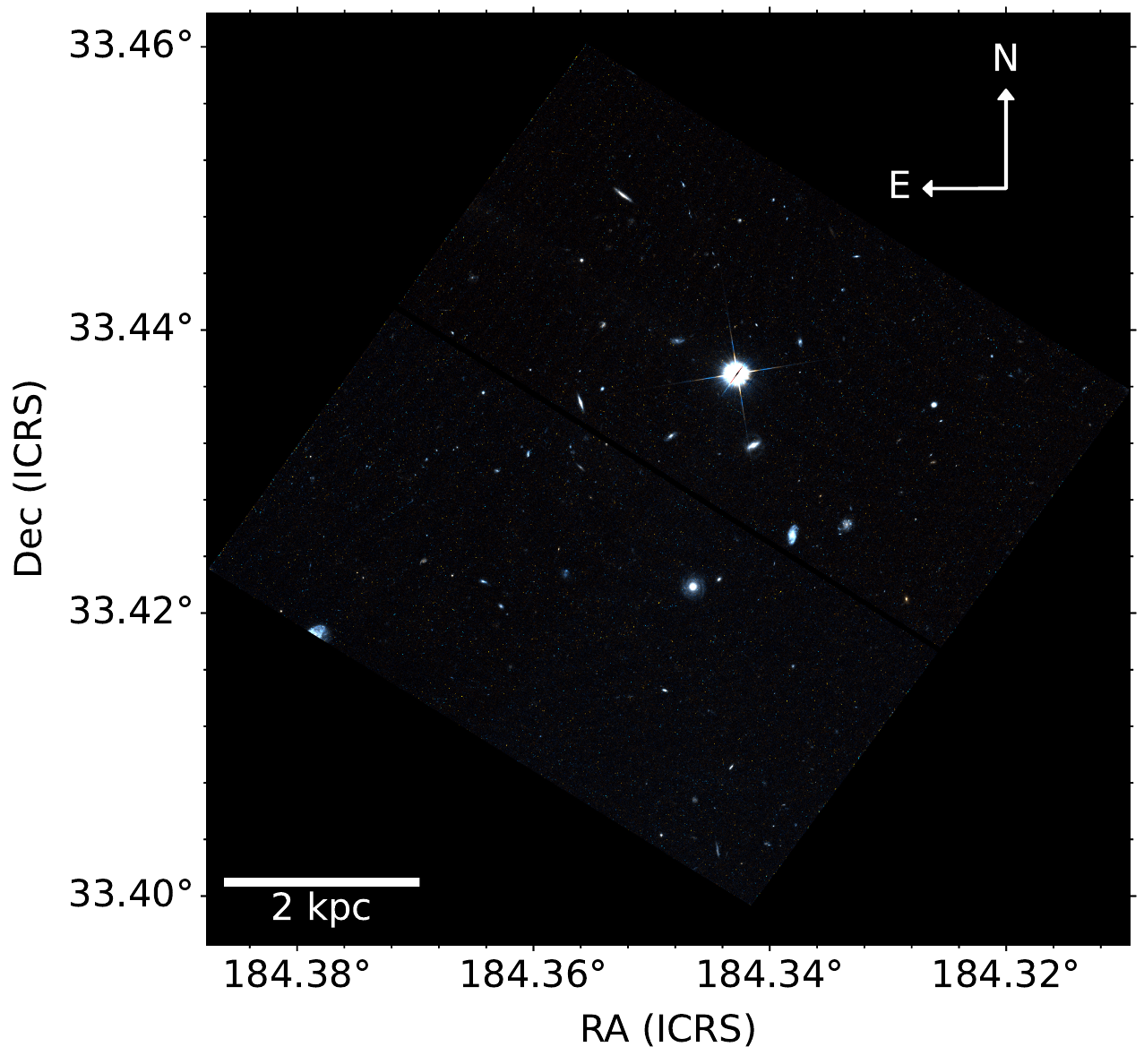}
    \end{subfigure}
    \caption{Color image of \CVnC{} based on the ACS observations (left) and the WFC3 parallel field (right). A white $2R_e$ ellipse on the ACS image based on our measured structural parameters shows the location of \CVnC{}. The F606W image for each instrument is colored blue, F814W is colored red, and the average of the F606W and F814W filters is shown in green. North is up and East is left. The image was made from the \texttt{flc} (charge transfer efficiency corrected) images of \CVnC{} using the HST \texttt{drizzlepac} and \texttt{APLpy} python packages \citep{Hack2013,Avila2015,aplpy2012,aplpy2019}.}
    \label{fig:three-color}
\end{figure*}

In Figure \ref{fig:three-color}, we present color images of the ACS observations (left) and WFC3 observations (right). A clear over-density of stars is seen in the image, which comprises the main stellar disk of \CVnC{}. \CVnC{} is largely confined to a single chip; both the second ACS chip as well as the parallel field show minimal foreground and background contamination.

\subsection{Photometry}
\label{sec:photometry}

We performed photometry on the \texttt{flc} images using the photometric software package \texttt{DOLPHOT} \citep{Dolphin2000,Dolphin2016}. Most \texttt{DOLPHOT} parameters were set according to those recommended in \citet{Williams2014,Williams2021} except for \texttt{CombineChi}, for which we chose to use the default value of \texttt{CombineChi=0}\footnote{We note that while this choice of CombineChi increases the number of detections passing our subsequent quality cuts, the signal-to-noise is likely overestimated in cases where bad detections should have been degraded in weight.}. We filtered the \texttt{DOLPHOT} catalog for well-recovered, high-fidelity sources by applying numerous quality cuts. We only accepted sources with object type $\leq 2$, signal-to-noise $\geq 4$ in both filters, $\text{sharpness}_{F606W}^2 + \text{sharpness}_{F814W}^2 < 0.075$, and $\text{crowding}_{F606W} + \text{crowding}_{F814W} < 0.1$. Sharpness is a measure of the width of the brightness profile of the source relative to the PSF, and is used to filter out cosmic rays (which have a more positive sharpness) and background sources (which have a more negative sharpness value). Crowding is a measure of how much the brightness of the source would change if nearby stars in the image had not been fit simultaneously.

\texttt{DOLPHOT} was also used to carry out Artificial Star Tests (ASTs), in which 500,000 artificial stars were injected into each image one by one (following the spatial distribution of actual sources in \CVnC{}) and photometrically recovered. The same quality cuts used for the \CVnC{} stellar catalog were applied to the AST output. From these tests, we calculated the recovery fraction (the ratio of the number of stars recovered by \texttt{DOLPHOT} and the number of input stars) in each filter. Based on the ASTs, we measure 50$\%$ completeness limits of 27.58 mag in the F606W filter and 26.77 mag in the F814W filter.

\section{Results}
\label{sec:results}
\subsection{Structural Parameters} \label{sec:struc_params}

In order to provide a robust spatial cut for our photometry before further analysis, we measured the structural parameters of \CVnC{}. This also allowed us to compare the galaxy's properties to those of other systems (See Section \ref{sec:lum_mass}).

\begin{figure}
    \includegraphics[width=1.03\linewidth, alt={Corner plot showing the posterior distributions of the structural parameters derived for \CVnC{}. All distributions are roughly Gaussian, although the histograms corresponding to the effective radius and position angle have slightly extended tails and the distribution of ellipticities is slightly skewed towards higher ellipticities.}]{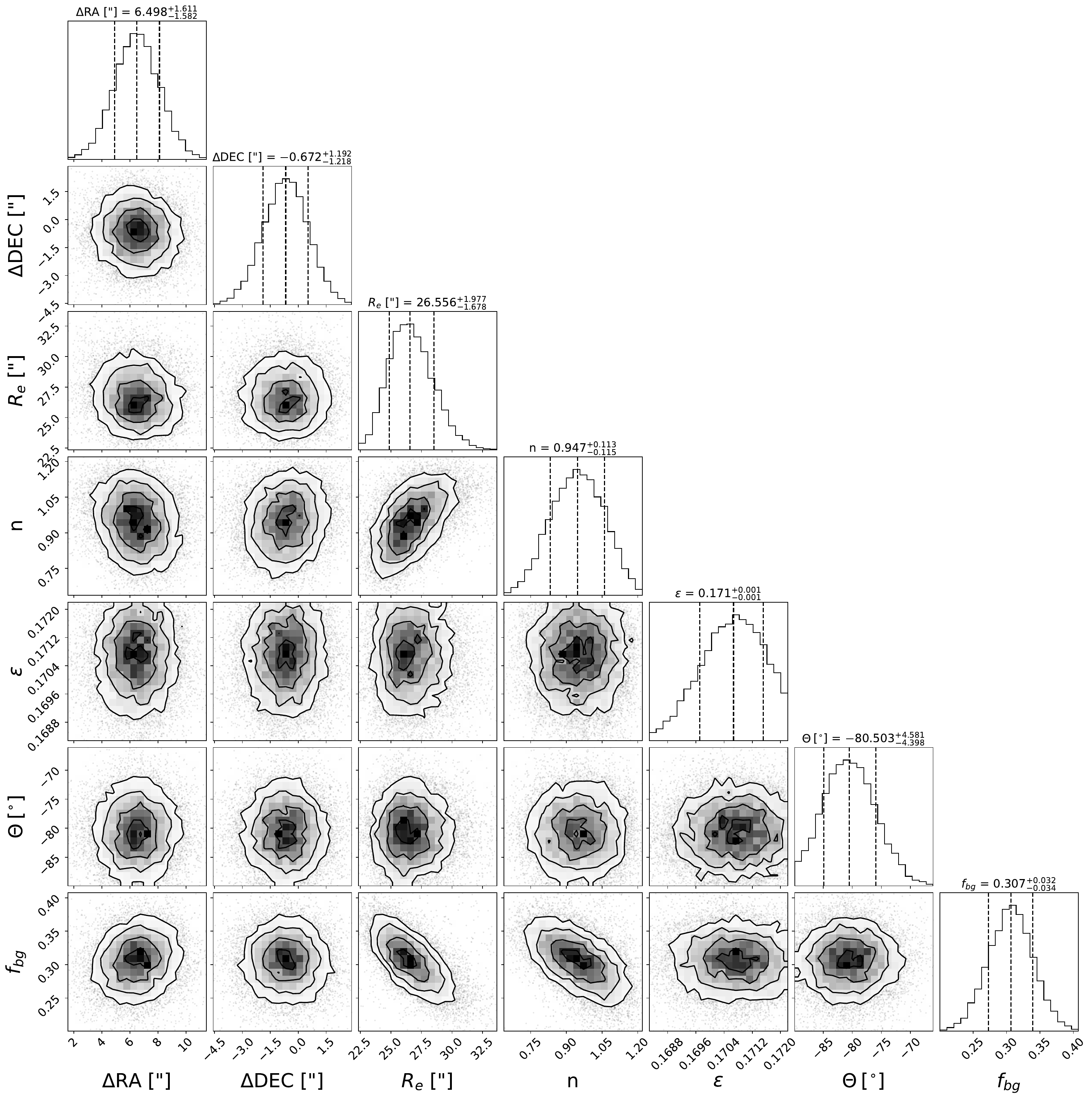}
    \caption{Posterior Distributions of our measured structural parameters of \CVnC{} based on fitting the spatial distribution of the resolved stars over 12,200 MCMC iterations. The $\Delta$RA and $\Delta$Dec values are the difference between the posterior distribution of \CVnC{}'s centroid and our initial by-eye estimate of the center of the distribution of well-recovered sources in \CVnC{}. Vertical dashed lines in each histogram show the best-fit values and $1\sigma$ errors, while contours in each scatter plot show 1, 2, and 3$\sigma$ errors. Best-fit values for all parameters are reported in Table 1.}
    \label{fig:struc_params}
\end{figure}

To estimate structural parameters, we do a Bayesian inference on the parameters for an elliptical S{\'e}rsic profile \citep{sersic}, fit to the sources that are likely to be stars following the criteria described in Section \ref{sec:photometry}. We assume a mixture model composed of a S{\'e}rsic profile and uniform background, with a free parameter $f_{\rm bg}$ setting the fraction of stars assigned to the background that is given a uniform prior between 0 and 1.  The S{\'e}rsic profile parameters are the half-light radius $R_e$ (along the major axis) and the S{\'e}rsic parameter $n$. For $R_e$ we use a scale-free prior (i.e., uniform in $\log(R_e)$; this is sometimes known, inaccurately, as a ``Jeffrey's'' prior) truncated at $4"$ and $60"$ (this truncation speeds convergence but has no significant impact on the posterior as Figure \ref{fig:struc_params} shows).  For $n$ we use a beta distribution prior with parameters $\alpha=6$, $\beta=3$, and an offset and rescaling of $0.5$ and $0.75$ (i.e., the support is $[0.5, 1.25]$).  This is in effect a prior that is roughly Gaussian, centered on 1, but with a longer tail to lower $n$, with the goal of allowing profiles roughly between exponential ($n=1$) and Gaussian ($n=0.5$), consistent with typical dwarf galaxies \citep{2018ApJ...860...66M}.  This prior is mildly informative in that the posterior (Figure \ref{fig:struc_params}, 4th row/column) is of roughly the same size as the prior, but the posterior is shifted from the peak and $\sim 10 \%$ narrower.   The location parameters of the profile are the coordinates of the center, parametrized as $\Delta {\rm RA}$ and $\Delta {\rm DEC}$, both of which are taken to be uniform distribution of $30"$ on either side of the by-eye estimated center. The shape parameters are position angle ($\Theta$; east form north) and the eccentricity ($\epsilon$).   For $\Theta$ we use a uniform prior from -90 to 90 degrees. The prior on $\epsilon$, however, is a strongly informative prior. It was taken to be a truncated Gaussian prior with $\mu=0.17$ and $\sigma=0.05$, truncated at 0 and 1. We came to the conclusion that an informative prior was needed by earlier attempts that used less informative priors, but where the ``best'' fit incorrectly fit the shape of the field-of-view instead of the galaxy itself. In practice, constraining any one of $R_e$, $\epsilon$, or $\Theta$ was sufficient to avoid this maximum in the likelihood, so we chose $\epsilon$ because it was most straightforward to estimate by-eye.  With this model and these priors in-hand, we inferred the parameters by adopting a likelihood that is simply the surface density of the stars, and fitting that to the model using the {\it dynesty} nested sampling code \citep{dynesty}.

Figure \ref{fig:struc_params} shows a corner plot with posterior distributions for each fitted parameter. Our final values for \CVnC{}'s sky position, half-light radius $R_e$, ellipticicty ($\epsilon=1-\frac{b}{a}$), position angle $\Theta$, S\'ersic index $n$, and fractional background density $f_{bg}$ are listed in Table \ref{tab:CVnC_prop}. We assume that the majority of the stellar populations of \CVnC{} detected in our data are within $2R_e$ (white ellipse in Figure \ref{fig:three-color})
and adopt this as the spatial boundary for our final high-fidelity photometric catalog of the galaxy.

\begin{figure}[ht!]
    \centering
    \includegraphics[width=1.0\linewidth, alt={Scatter plot (Declination vs.\ Right Ascension) showing the distribution of high-fidelity sources in the ACS field, which are highly concentrated in the central $2R_e$ region of \CVnC{}. A more sparse distribution of sources inhabit the region outside $2R_e$ of \CVnC{}'s center.}]{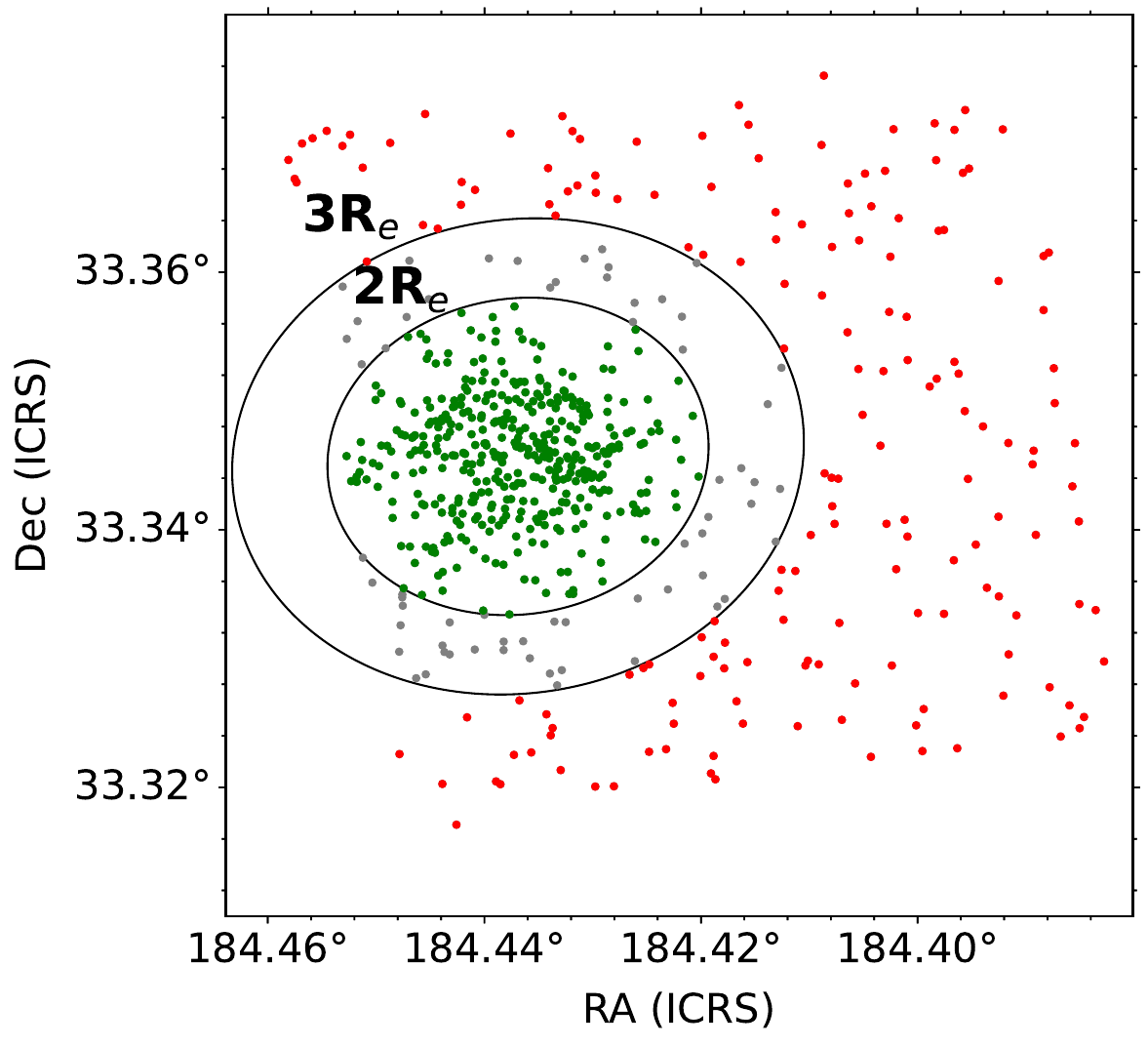}
    \caption{The spatial distribution of sources passing our quality cuts. The overplotted ellipses are based on the structural parameters of \CVnC{}. The inner $2R_e$ ellipse encompasses sources (shown in green) which are high-probability members of \CVnC{} and are used for further analysis of the system. Stars outside the outer $3R_e$ ellipse (shown in red) are used as a measure of contamination. Sources in between $2-3\:R_e$, shown in grey, were not used in either analysis.}
    \label{fig:spatial_cmd_cuts}
\end{figure}

In Figure \ref{fig:spatial_cmd_cuts}, we overlay ellipses based on our final geometry with a semi-major axis of 2$R_e$ centered on the spatial distribution of sources to show the extent of our measured structural parameters. To quantify the potential background contamination of the \CVnC{} CMD, we conservatively use sources in the off-target ACS field, recovered outside $3R_e$ of the galaxy (shown in red in Figure \ref{fig:spatial_cmd_cuts}). We choose to use our ACS off-target ($>3R_e$) field for the measurement of the background density instead of the WFC3 parallel imaging, since the WFC3 field is underpopulated and the ACS field thus provides a more conservative estimate of the background.

\setlength{\tabcolsep}{15pt}
\hspace{-15pt}
\begin{table}[h]
\caption{\CVnC{} Properties}
    \begin{tabular}{lr}
        \hline
        \hline
        Property & Value\\
        \hline
        RA ($\degree$ J2000) & $184.4369\pm0.0004$\\
        Dec ($\degree$ J2000) & $33.3457^{+0.0003}_{-0.0003}$ \\
        P.A. $\Theta$ ($\degree$ E of N) & $-81^{+5}_{-4}$\\
        ellipticity ($\epsilon=1-\frac{b}{a}$) & $0.1707^{+0.0008}_{-0.0010}$\\
        S\'ersic $n$ parameter & $0.94^{+0.11}_{-0.12}$\\
        $f_{bg}$ & $0.307^{+0.032}_{-0.034}$\\
        $R_e$ ($\arcsec$) & $26.6^{+2.0}_{-1.7}$\\
        $R_e$ (kpc) & $1.09^{+0.10}_{-0.08}$\\
        $\rm M_V$ (mag) & $-11.2^{+0.8}_{-0.5}$\\
        $\rm M_*$ ($\rm M_{\odot}$) & $3.4^{+4.2}_{-2.6}\times10^6$\\
        $\rm M_{\text{H}\textsc{i}}$ ($\rm M_{\odot}$) &$<1.5\times10^6$\\
        log$(sSFR$ ($yr^{-1}$)) &$-13.2^{+0.6}_{-0.4}$\\
        $\tau_{90}$ (Gyr) & $6.5^{+0.1}_{-5.5}$\\
        TRGB $\rm m_{\I}$ (mag) & $25.58^{+0.12}_{-0.08}$\\
        $\mu$ (mag) & $29.63^{+0.12}_{-0.08}$\\
        Distance (Mpc) & $8.43^{+0.47}_{-0.32}$\\
        $D_{\text{NGC~4631}}$ (Mpc) & 1.29\\
        $\Theta_{5}$ & $-$0.20\\
        $\text{[M/H]}$ (dex) & $-0.6$\\
        $A_V$ (mag) & 0.035\\
        $A_{\V}$ (mag) & 0.034\\
        $A_{\I}$ (mag) & 0.021\\
        F606W 50\% (mag) & 27.58\\
        F814W 50\% (mag) & 26.77\\
        \hline
        \hline
        \label{tab:CVnC_prop}

    \end{tabular}
    {\raggedright NOTE---These are the basic properties of \CVnC{} measured in this work, with the exception of forgeround extinction corrections from \citet{Schlegel} and \citet{schlafly}, the H\textsc{i} mass upper limit as measured using data from \citet{Nazarova2025}, and the sSFR as measured using FUV magnitudes from \citet{Karachentsev2013}. $\mu$ refers to the distance modulus of \CVnC{}, while $D_{\text{NGC~4631}}$ is the system's distance to its most influential $L_{\star}$ neighbor, NGC~4631. $\Theta_5$ is \CVnC{}'s tidal index based on the five neighboring galaxies with the most tidal influence. The 50\% magnitudes are 50\% completeness limits for each filter measured using ASTs. \par}
\end{table}

\begin{figure}
    \centering
    \includegraphics[width=1.0\linewidth, alt={Scatter plot of the binned observed density of stars in \CVnC{} as a function of radius from the center of the galaxy. An overplotted curve shows the profile corresponding to our derived structural parameters, which lines up with the observed densities for all but 4 of the 28 points, within error bars.}]{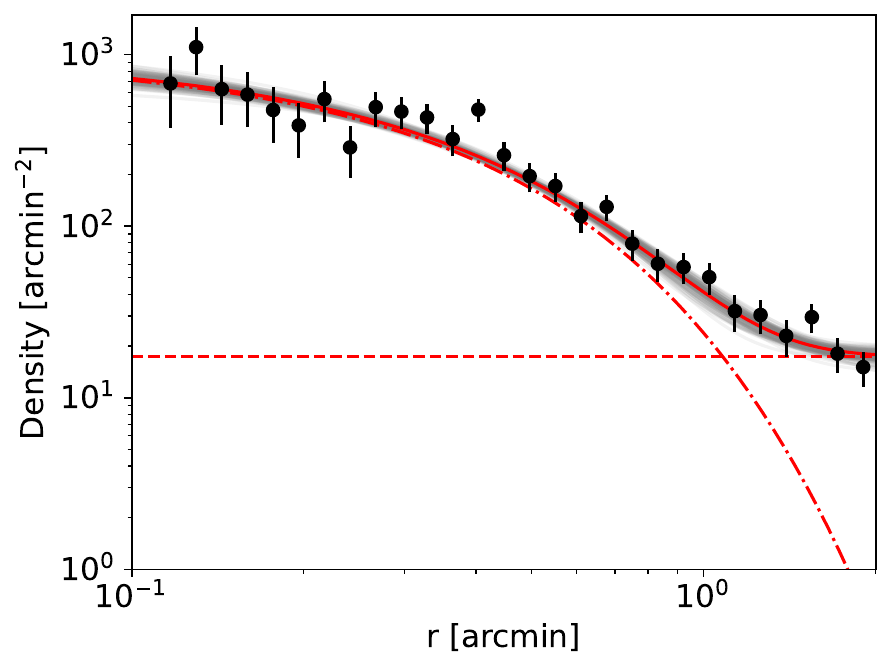}
    \caption{The radial density profile of \CVnC{}. A dot-dashed red line shows the S{\'e}rsic profile corresponding to the median structural parameters determined from our Bayesian inference approach. The red horizontal dashed line shows the uniform background density corresponding to the best-fitting $f_{bg}$, and the solid red line shows the sum of the S{\'e}rsic profile and uniform background. Grey lines show the same for 100 sets of parameters randomly sampled from the posteriors of our fit. Black points show the observed stellar density profile of \CVnC{} and error bars show the corresponding Poissonian uncertainties. The observed stellar densities at large radii have been corrected for fraction of the corresponding area falling outside the ACS field of view.}
    \label{fig:rad_prof}
\end{figure}

In Figure \ref{fig:rad_prof}, we check our structural parameters by showing the observed binned radial density profile of \CVnC{} (black points) overplotted on the profile expected from our inferred parameters (red solid line). Grey lines show similar profiles from 100 random sets of structural parameters sampled from the posterior distributions of our Bayesian inference approach. Dash-dotted and dashed lines show the S{\'e}rsic profile and uniform background component of the fit, respectively. Observed stellar densities at outer radii ($>1$ arcmin) have been corrected for the portion of the corresponding area falling outside the ACS field of view. The observed profile is in excellent agreement with that predicted from our structural parameters. 

We also verify that the inner regions of \CVnC{} are unaffected by crowding, which otherwise may influence the reliability of our structural parameters. We measure the average separation of the 50 innermost stars in \CVnC{} to be $\sim10\arcsec$, which is much larger than the size of the ACS/WFC Point Spread Function (PSF) at the studied wavelengths. Thus, we conclude that our imaging of \CVnC{} is not crowding-limited at the studied depth and wavelength coverage of our imaging.

\subsection{Color-Magnitude Diagram} \label{sec:cmd}

\begin{figure*}[ht]
    \begin{subfigure}[t]{0.72\textwidth}
    \includegraphics[width=1.0\linewidth, alt={The first two panels show the color-magnitude distribution of stars in \CVnC{} as a scatter plot and a binned scatter plot (covering a 7366 square arsec region); both show clear overdensities of stars in the RGB region, with about 18 out of 401 stars above or blue-ward of the RGB. The third plot, which shows the binned CMD of stars in the field ACS region (26309 square arsec), shows an absence of a clear RGB and a few background/foreground objects above and blue-ward of the RGB. The fourth panel shows the CMD of sources in \CVnC{} again, this time colored by the probability of being a contaminant, and show that these blue and/or brighter sources are likely contaminants. However, there are also a few sources closer to the TRGB which may be real \CVnC{} stars.}]{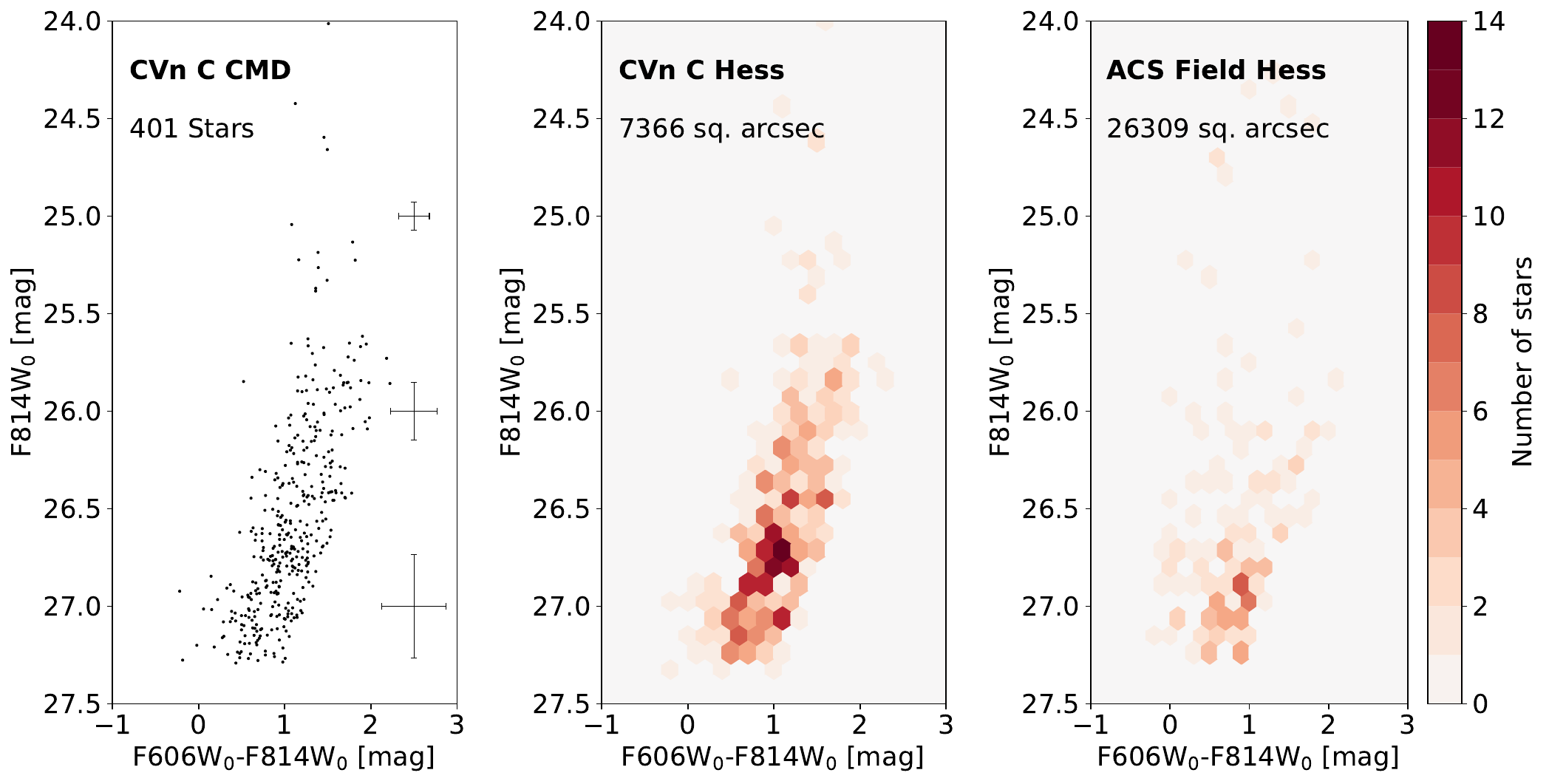}
    \end{subfigure}
    \hspace{0pt}
    \begin{subfigure}[t]{0.28\textwidth}
    \includegraphics[width=1.0\linewidth]{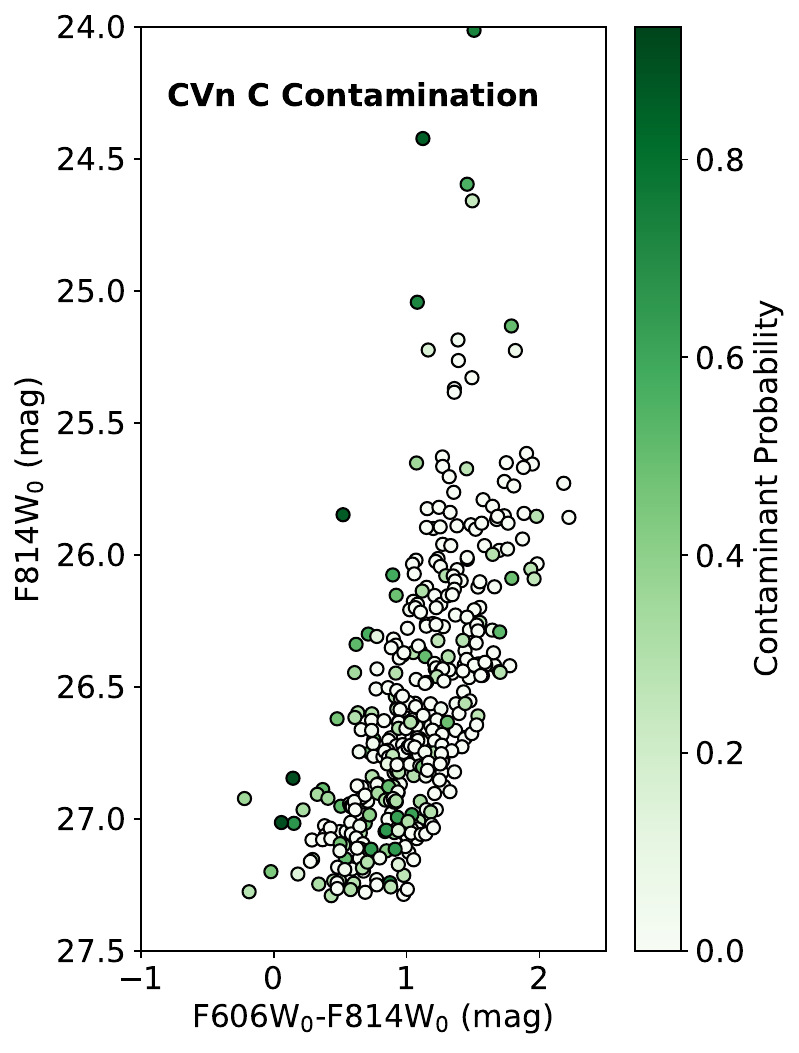}
    \end{subfigure}
    \caption{First Panel: CMD showing the 401 stars passing quality and spatial cuts in \CVnC{}, with error bars showing the RMSE in three magnitude bins. The data points have been corrected for Galactic extinction based on the values listed in Table \ref{tab:CVnC_prop}. Second Panel: Hess diagram showing the binned density of stars in \CVnC{}. Third Panel: Hess diagram showing the binned density of stars in the off-target ACS Field region used to identify background contamination. Both  binned density diagrams are annotated with the area of the corresponding region. \CVnC{}'s galaxy region shows a clear RGB which is absent from the background field. 
    Fourth Panel: Same CMD as first panel, now colored by the probability of the source being a contaminant. The probability is measured based on the proximity of the source in color-magnitude space to sources in the off-target ACS field region, as described in Section \ref{sec:cmd}. Sources brighter than or blue-ward of the RGB have a high probability of being foreground or background contaminants.}
    \label{fig:field_hex}
\end{figure*}

The left-most panel in Figure \ref{fig:field_hex} shows the extinction-corrected CMD of the 401 sources in \CVnC{} which pass the quality cuts and lie within 2$R_e$ of the galaxy center based on our structural parameters. Error bars show the uncertainties in three representative magnitude bins. These uncertainties correspond to the root mean square error (RMSE) in each bin, estimated using $\rm m_{out}-m_{in}$ values from ASTs. The second panel shows a Hess diagram of these sources, where darker colors signify a higher density of sources. The third panel shows a similar Hess diagram of sources in the background field region of the ACS imaging ($>3R_e$, shown spatially in Figure \ref{fig:spatial_cmd_cuts}) which passed the same quality cuts applied to the sources in \CVnC{}. Note that each Hess diagram pulls from distinctly different areas; the area of the region encompassing the relevant sources is annotated in the panels. As shown in Figure \ref{fig:field_hex}, \CVnC{} has a clear Red Giant Branch (RGB) structure which is absent in the background region. 

There are also some sources in the \CVnC{} CMD which sit outside of the RGB region, including a few points above the RGB ($\rm m_{F814W}\sim 25.5$ mag) and a smaller number of sources blue-ward of the RGB, both at faint magnitudes ($\rm m_{F814W}\sim 26.8-27.3$ mag) and near the TRGB ($\rm m_{F814W}\sim 25.7$ mag). Upon comparison to the Field Hess diagram in Figure \ref{fig:field_hex}, a number of these sources match the distribution of those in the off-target ACS field and thus may be foreground or background sources. 

To assess the membership of each source to the target dwarf galaxy, we use a Monte Carlo (MC) approach to assign a probability to each source of being a contaminant based on their position in the CMD relative to the off-target field CMD\footnote{We verify that the completeness limits of the data are similar in the galaxy region compared to the off-target field. Hence, we do not make completeness considerations when comparing the color-magnitude distribution of sources between the two regions.}. We first calculate the expected number of contaminant sources in the \CVnC{} ellipse ($<2R_e$) by scaling the number of sources in the off-target ACS field ($>3R_e$) by the relative areas encompassed by the two regions, i.e., $(\text{number of field sources})\times\frac{\text{galaxy ellipse area}}{\text{off-target ACS field area}}$. Based on this calculation, we expect $\sim48$ contaminant sources in our CMD for \CVnC{}. We then randomly select a subsample of 48 sources from the off-target ACS field region and assign a Nearest Neighbor (NN) in color-magnitude space in the galaxy region to each field source. The NN is chosen to be the source in the galaxy $2R_e$ ellipse with the shortest weighted Euclidean distance to the field source, where the weight corrects for the unequal baselines in color and magnitude. This process is repeated for 1000 iterations, randomly choosing sub-samples of the field source population. The fraction of times each source in \CVnC{}'s CMD is assigned as a NN is then chosen as the probability of the source being a contaminant, since it represents the proximity in color-magnitude space of the source to background/foreground sources in the field region. We use the ACS off-target field to measure this foreground/background contamination instead of the WFC3 field because, as noted in Section \ref{sec:struc_params}, our WFC3 parallel field is underpopulated.

The fourth panel of Figure \ref{fig:field_hex} shows the sources in the \CVnC{} CMD, now colored by the probability of being a contaminant. The brightest sources above the RGB and all sources blue-ward of the RGB have a high probability of being a contaminant. Indeed, upon visual inspection of the color image, the brightest sources were found to be likely foreground stars or associated with foreground stars.

A few sources  $\sim0.5$ mag above the TRGB ($\rm m_{F814W}\sim 25.2-25.5$ mag) appear to be legitimate stars in \CVnC{}, both in terms of their low contaminant probability and after visual inspection. We find that the location of these stars in color-magnitude space is consistent with the AGB phase of 1 Gyr isochrones, and thus we conclude that these are likely to be intermediate-age AGB stars.

To determine the effects of the inclusion of contaminants on our downstream analysis and results, we remeasure the TRGB distance, SFH, and environment around \CVnC{} with a `clean' CMD. The `clean' CMD comprises only of sources which have a $>0.65$ probability of not being a contaminant, where the probability threshold is chosen such that the number of sources excluded equals the number of sources expected to be contaminants in the galaxy ellipse region. We find that the TRGB distance, SFH, and present-day stellar mass derived from the SFH are all unchanged within uncertainties. The environment of \CVnC{} (see Section \ref{sec:envir}) also remains unchanged, with only one more (dwarf) galaxy neighbor found within 1.5 Mpc of \CVnC{}. Due to the robustness of our results to the inclusion of these likely contaminants and for the sake of statistical rigor, we choose to include all sources that meet our quality cuts and lie within our spatially selected region in our final catalog and downstream analysis.

\subsection{TRGB Distance Determination} \label{sec:trgb}

We determine a TRGB distance to \CVnC{} of $8.43^{+0.47}_{-0.32}$ Mpc based on the system's CMD, which shows a defined RGB and a clearly identifiable TRGB.
The absolute magnitude of the TRGB in the I-band is known to be remarkably constant across stars with different ages and metallicities; while there is a modest dependency on the luminosity of TRGB stars with metallicity in the I-band, this dependence is well-measured and accounted for in TRGB calibrations. Thus, the apparent magnitude of the TRGB is an accurate and precise indicator of distance to a galaxy.

We determined the TRGB magnitude of \CVnC{} in the F814W filter using the approach described in detail in \citet{newman} and briefly summarized here. We first corrected the \CVnC{} CMD for foreground extinction using dust maps from \citet{Schlegel} and the re-calibration from \citet{schlafly}. We correct for the TRGB's small dependence on metallicity using a Quadratic Transformation (QT) color correction from \citet{Jang_2017}, using Equation \ref{eq:color_corr} which we reproduce here for reference:

\begin{equation}
\label{eq:color_corr}
\begin{split}
    \I_{\mathrm{rect}}=\I&-\alpha[(\V-\I)-1.1]^2 \\
        &-
        \beta[(\V-\I)-1.1],\\
\end{split}
\end{equation}

\noindent where $\I_{\mathrm{rect}}$ is the corrected magnitude and $\alpha=0.159\pm0.010$ and $\beta=-0.047\pm0.020$ are best-fit parameters from \citet{Jang_2017} which describe the shape of the color dependence. This correction was only made for sources with color $(\V-\I)>1.5$ mag, because the TRGB luminosity is noted to be approximately constant at colors blue-ward of $V-I\sim 1.5$ mag.

After correcting the CMD for foreground extinction and color-dependency, we employed the Gaussian-weighted Locally Estimated Scatter plot Smoothing (GLOESS) algorithm, which has been applied both to TRGB measurements \citep[e.g.,][]{Hatt2017,newman} as well as to Cepheid and RR-Lyrae light curves \citep[e.g.,][]{Persson2004,Monson2017}. GLOESS is a non-parametric method which uses a least-squares algorithm to fit the observed stellar luminosity function (LF) with Gaussian weighting and a user-defined smoothing scale, $\tau$. A key advantage of GLOESS is that the smoothing scale can be adjusted to minimize the contribution of false edges (relative to true edges) in the LF, which is especially important for sparse LFs such as \CVnC{}'s. We chose a smoothing scale $\tau = 0.11$ and a binwidth of 0.01 to smooth the LF, after which we applied a Sobel
[+1,0,-1] edge detection kernel and chose the point of greatest response in the LF as the TRGB.

To estimate the uncertainties in our measurement, we first used the ASTs measured for \CVnC{} (see Section \ref{sec:photometry}) to calculate statistical uncertainties for each point in the observed LF. We found the 1000 ASTs with the shortest Euclidean distance in color-magnitude space to each real observation in \CVnC{}'s stellar catalog, and recorded the average difference between the recovered and input F814W magnitudes ($\rm m_{out}-m_{in}$) of the 1000 ASTs; this became the error for that observation. We subsequently sampled 5000 random estimates of the F814W magnitude for each point source in the CMD, using a normal distribution centered on the observed magnitude with a standard deviation equal to the corresponding error in that measurement. To minimize the detection of false edges, we chose sources between the magnitude limits corresponding to the approximate luminosity of the TRGB ($\rm m_{F814W}=25.0-26.0$). We subsequently made 5000 independent measurements of the TRGB using the GLOESS method with each of our independent samples of magnitude estimates. We choose the median of the 5000 measurements as our final TRGB and estimate our $1\sigma$ errors based on the 16th and 84th percentiles. Using this approach, we obtained a TRGB apparent magnitude of $\I_{\T}=25.58^{+0.12}_{-0.08}\:$mag.

\begin{figure}[ht!]
    \includegraphics[width=1.0\linewidth, alt={A scatter plot showing the CMD of \CVnC{}, accompanied by a sideways histogram of the F814W magnitude of sources in \CVnC{}. A horizontal line shows the location of the TRGB (25.58 mag), which lines up with the drop of stellar densities in the histogram. The uncertainties in the TRGB magnitude are plotted, which span the 25.5-25.7 mag range. 10 Gyr isochrones with different metallicties are overplotted and match the rough color-magnitude distribution of the RGB.}]{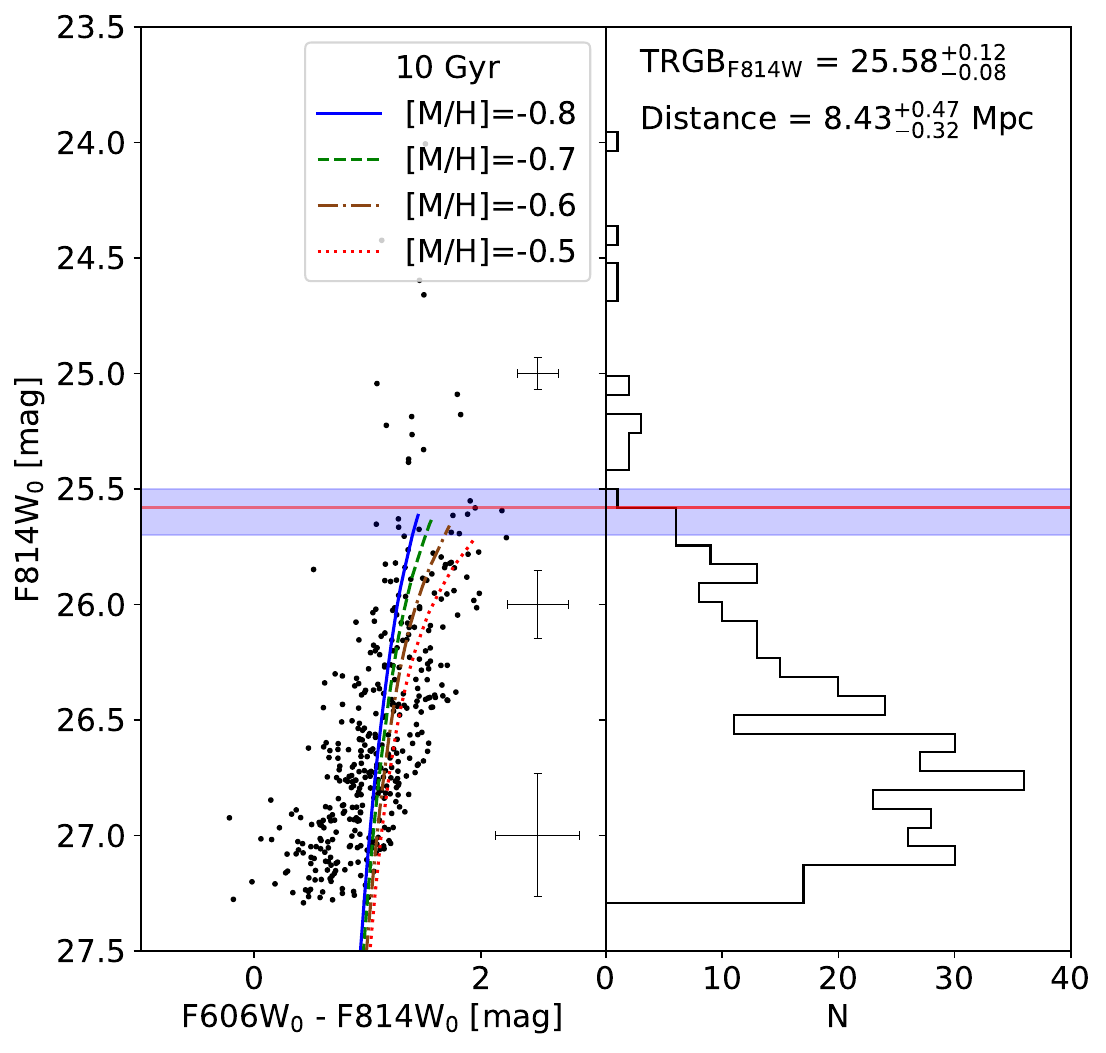}
    \caption{Left: CMD of \CVnC{} after applying the color-based metallicity correction for TRGB fitting, with error bars showing mean uncertainties in three magnitude bins. Right: the stellar luminosity function of \CVnC{} based on the CMD shown. The red horizontal line represents the best-fitting TRGB magnitude for the galaxy, while the purple shaded region shows the associated statistical uncertainties in the fit. Colored lines in the left plot show 10 Gyr PARSEC \citep{Bressan2012} isochrones with metallicities of $[\rm{M/H}]=-0.5$ (red dotted line), $[\rm{M/H}]=-0.6$ (brown dash-dotted line), $[\rm{M/H}]=-0.7$ (green dashed line), and $[\rm{M/H}]=-0.8$ (blue solid line).}
    \label{fig:cmd}
\end{figure}

To obtain a distance modulus from $\I_{\T}$, we used the F814W TRGB zeropoint provided in \citet{Freedman_2021}, $\mathrm{M}_{\T}=-4.049\pm0.015\:\text{(stat)}\pm 0.035\:\text{(sys)}$ mag. We calculate a distance modulus of $29.63^{+0.12}_{-0.08}\:\text{(stat)}\pm0.04\:\text{(sys)}$ mag. The statistical uncertainties in the distance modulus include the uncertainties from the TRGB magnitude and the statistical uncertainty in the F814W zeropoint (0.015 mag). The systematic uncertainties include the uncertainty in the extinction correction (which we take to be 10\% of the $A_{F814W}$, or 0.002 mag), the uncertainty in the QT color-correction (estimated to be 0.01 mag), and the systematic uncertainty in the F814W zeropoint (0.035 mag). This converts to a distance of $8.43^{+0.47}_{-0.32}\:\text{(stat)}\pm0.14\:\text{(sys)}$ Mpc, which will hereafter be quoted only with the statistical uncertainties since they dominate over the systematic errors. Figure \ref{fig:cmd} presents our final color- and extinction-corrected CMD of \CVnC{}, with the TRGB magnitude indicated in red and the statistical error in the TRGB magnitude indicated in purple. \CVnC{} previously had a redshift-based distance of 4.7 Mpc \citep{Karachentsev2018}; this new TRGB distance puts the galaxy at almost twice its original recorded distance.

To estimate \CVnC{}'s metallicity, we use 10 Gyr PARSEC isochrones \citep{Bressan2012} shifted to the distance modulus derived above, and examine by-eye which fit best on the CMD. Figure \ref{fig:cmd} is overplotted with the four best-fitting isochrones, from which we estimate a representative metallicity of $[\rm{M/H}]=-0.6$. This metallicity also agrees well with the range reported by our SFH-fitting using MATCH (Section \ref{sec:sfh_match}).

\subsection{The Star Formation and Gas Content of Canes Venatici C} \label{sec:sfh}

Here we describe three independent probes of the star formation in \CVnC{}: constraints on the system's recent star formation using CMD-fitting (Section \ref{sec:sfh_match}), measurement of \CVnC{}'s recent ($\sim100$ Myr) star formation rate (SFR) using archival FUV data (Section \ref{sec:sfh_fuv}), and an upper limit on \CVnC{}'s neutral hydrogen (H\textsc{i}) reservoir using archival H\textsc{i} surveys (Section \ref{sec:gas_content}). Constraints from all three of these independent probes suggest that \CVnC{} is a quenched, gas-poor galaxy with little---if any---star formation over the past few 100 Myr.

\subsubsection{Star Formation Constraints from CMD-fitting}\label{sec:sfh_match}

CMDs can be fit to extract a SFH, i.e., the variation of the SFR of a galaxy with time SFR(t). This is a powerful approach for constraining star formation in nearby dwarf galaxies \citep[e.g.,][]{Tolstoy2009,McQuinn2009,McQuinn2010,brown2014,Weisz2014a,Weisz2014b,Sacchi2021,McQuinn2023,Savino2023,McQuinn2024LeoMK}. However, reliable SFHs at early look-back times require a well-populated old Main Sequence Turnoff (oMSTO) \citep[e.g.,][]{Tolstoy2009} in the CMD. At the distance of \CVnC{}, the completeness limit of our data (see Table \ref{tab:CVnC_prop}) is several magnitudes brighter than the oMSTO, inhibiting a reliable estimate of the galaxy SFH at early look-back times. However, the depth of our current data does allow us to potentially constrain whether the galaxy has experienced any appreciable star formation in the last several 100 Myr.

We derived the SFH of \CVnC{} using the SFH-fitting code MATCH \citep{Dolphin2002}. MATCH uses stellar evolution libraries covering a range of ages and metallicities to create synthetic CMDs of galaxies for a given Initial Mass Function (IMF) and binary fraction. The user supplies photometric catalogs of stars and ASTs and MATCH then uses a Poisson likelihood statistic to find the synthetic CMD which best fits the observed CMD, and uses the underlying combination of synthetic simple stellar population (SSP) models to construct a best-fitting SFH.

To fit the SFH of \CVnC{}, we used a Kroupa IMF \citep{Kroupa2001} and a binary fraction of 0.35. We used the TRGB-derived distance modulus of 29.63 mag and a foreground extinction of 0.035 mag in the Landolt V band, calculated using using dust maps from \citet{Schlegel} and the re-calibration from \citet{schlafly}. We constrained the metallicity [M/H] to increase monotonically with age with a resolution of 0.15 dex.
The time binning in our SFH fits consists of 5 bins for the log(Age)$=8.00-9.00$ age range and two bins between log(Age)$=9.00-10.15$, following conventions in the literature for SFHs derived from shallower HST data \citep[e.g.,][]{Dolphin2001}. Due to a lack of recent star formation, we chose to fit the log(Age)$=6.60-8.00$ time period with a single bin; we verified that this choice does not change the SFH within uncertainties. We use both the PARSEC \citep{Bressan2012} and MIST \citep{Choi2016} evolutionary libraries for this fitting, which helps demonstrate the differences in fits based on the stellar models. We estimate statistical errors in the fit using a hybrid Markov Chain Monte Carlo approach \citep{Dolphin2013}. Systematic errors are estimated by refitting the SFH for 50 iterations with shifts in luminosity and color \citep{Dolphin2012}.

\begin{figure}
    \centering
    \includegraphics[width=1.0\linewidth, alt={Line plot showing the cumulative SFH of \CVnC{} as derived from two stellar evolution libraries. The SFHs from the libaries have negligible differences and match each other completely within the associated error envelopes. Both SFHs show no star formation for the last few 100 Myr, a small amount of star formation in the last Gyr (upto 10\% of the total mass of the galaxy), and quenching between $\sim$1-7 Gyr ago.}]{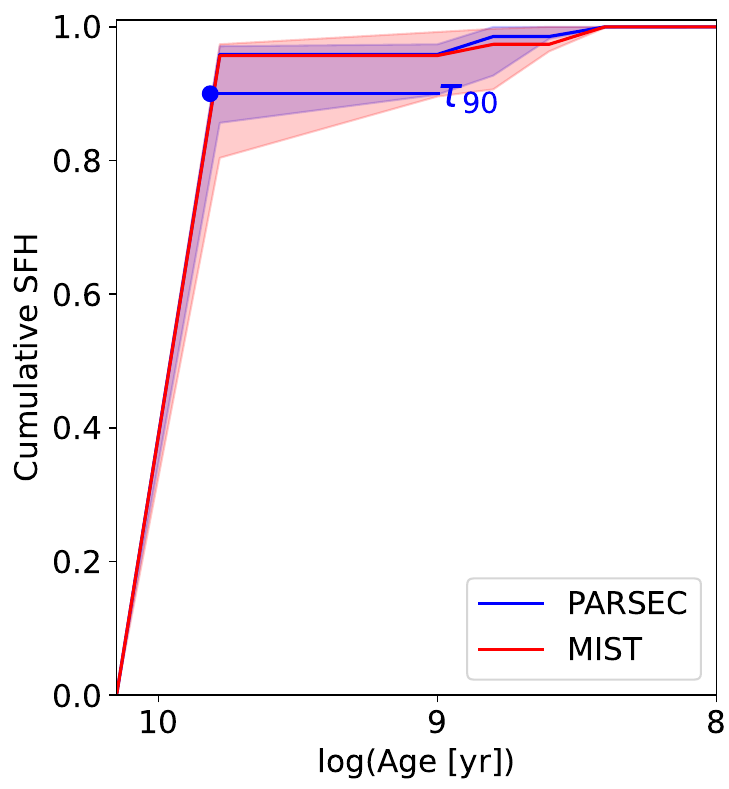}
    \caption{Best-fit Cumulative SFH, i.e., SFR(t) as a function of logarithmic lookback time and redshift, from CMD fitting using MATCH \citep{Dolphin2002}. The CSFH derived using PARSEC is shown in blue, while MIST is presented in red. The quenching timescale $\tau_{90}$ derived from PARSEC is overplotted with a blue line. Shaded regions show both systematic and statistical uncertainties. Fits from both libraries agree well and show early quenching $>1$ Gyr ago. For visualization purposes, we trim the x-axis since there is no recent ($<100$ Myr) star formation reported in the best-fit CSFH.}
    \label{fig:csfh}
\end{figure}

Figure \ref{fig:csfh} shows the best-fit cumulative SFHs (CSFHs) using the PARSEC and MIST libraries. The fits from both libraries agree well within uncertainties, so we choose to use only the PARSEC fit for the downstream analysis. The timescale $\tau_{90}$ at which \CVnC{} quenched, i.e., formed $90\%$ of its stars, is shown with a blue line which encompasses the error in our measurement. Although the CSFH has large uncertainties at early and intermediate ages, our results show little to no recent star formation and suggest the galaxy was quenched $6.5^{+0.1}_{-5.5}$ Gyr ago.

\begin{figure}
    \centering
    \includegraphics[width=1.0\linewidth, alt={Scatter plot showing \CVnC{}'s CMD, with isochrones of several ages overplotted as curves. A few stars blue-ward of the RGB and significantly brighter than the RGB line up with the younger (100 Myr to 1 Gyr) isochrones. Some of these have a contamination probability of more than $65\%$ (See Section \ref{sec:cmd}), making them likelike foreground or background sources. Most stars are red-ward of the 1 Gyr isochrone, indicating older ages.}]{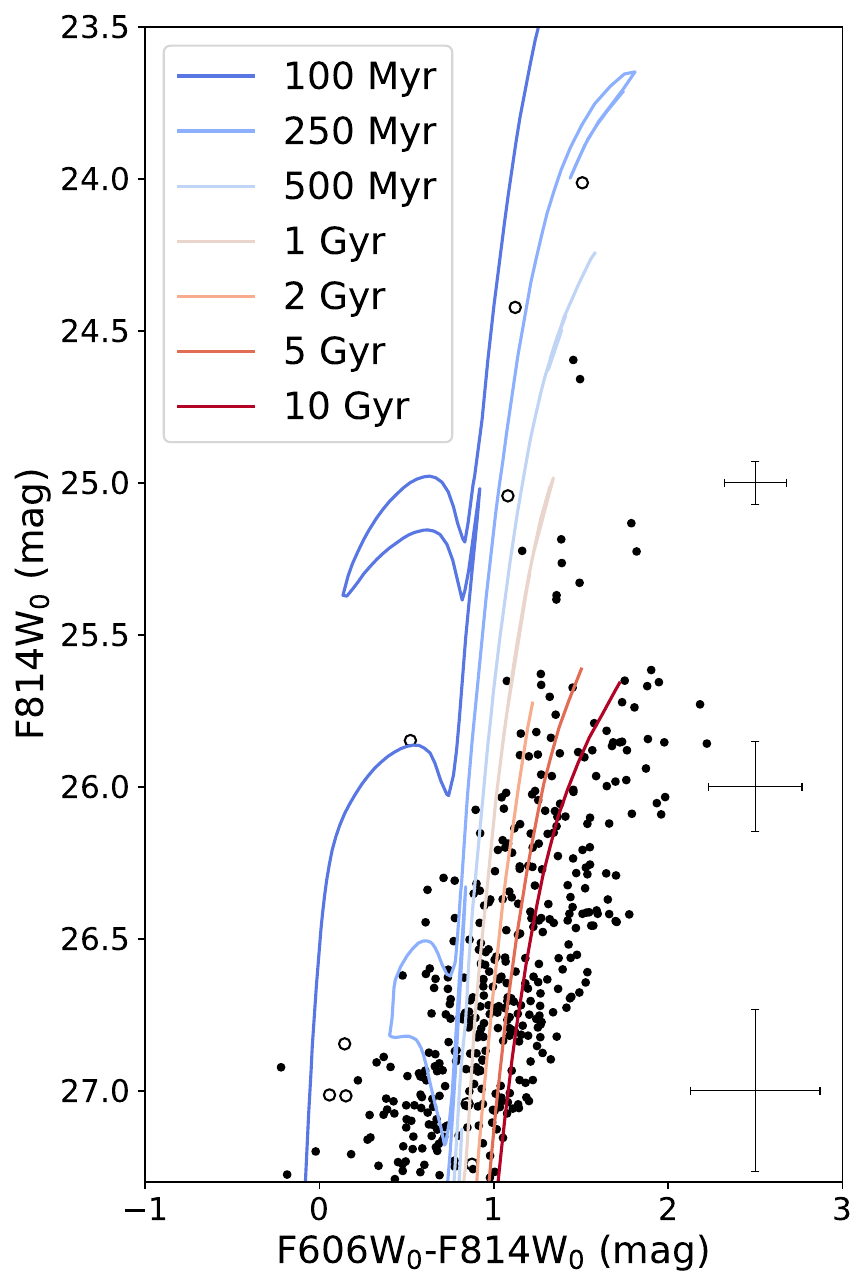}
    \caption{Extinction-corrected CMD of stars in \CVnC{}. Overplotted are isochrones with ages of (from left to right) 100 Myr, 250 Myr, 500 Myr, 1 Gyr, 2 Gyr, 5 Gyr, and 10 Gyr and a metallicity of [M/H]=$-$0.6 from the PARSEC library \citep{Bressan2012}.
    Open circles show the sources which were found to have a high ($>65\%$) probability of being a contaminant (see Section \ref{sec:cmd}). The majority of \CVnC{}'s stars are consistent with ages $>1$ Gyr.
    However, we are unable to rule out a small amount of late-time ($>100$ Myr) star formation.
    }
    \label{fig:iso_cmd}
\end{figure}

These results are demonstrated further by Figure \ref{fig:iso_cmd}, where we show the CMD for \CVnC{} now overplotted with PARSEC isochrones ranging from 100 Myr-10 Gyr \citep{Bressan2012}. Nearly all stars in the CMD are consistent with isochrones older than 1 Gyr. The exception to this are a few sources that sit bright-ward of the TRGB and whose photometric properties are consistent with ages between 100 Myr to 1 Gyr. Section \ref{sec:cmd} uses a MC approach to show that the brightest of these are likely foreground or background contaminants. These sources (i.e., all sources with a contamination probability of more than $65\%$) are shown as open circles. However, a few of these sources are consistent with being part of \CVnC{} and may be intermediate-age AGB stars. There are no sources in the CMD that are consistent with isochrones with ages less than 100 Myr. Hence, although \CVnC{} appears to be largely dominated by an old stellar population and a lack of young ($<100$ Myr) stars, a small amount of late-time star formation ($>100$ Myr) cannot be ruled out.

\subsubsection{Star Formation Constraints from archival FUV data}\label{sec:sfh_fuv}

Far-UV emission is an excellent tracer of recent ($<100$ Myr) star formation in nearby galaxies \citep[e.g.,][]{Lee2011,McQuinn2015}. For an independent probe of the recent star formation in \CVnC{}, we used its apparent FUV magnitude $\rm m_{FUV}$ to obtain a SFR for the system. The FUV archival imaging is from the Galaxy Evolution Explorer (GALEX; \citealp{Martin2005,Gil2007}), and the integrated flux for CVnC was reported in the Updated Nearby Galaxy Catalog (UNGC; \citealp{Karachentsev2013}). We used the FUV magnitude from the UNGC ($\rm m_{FUV}=22.80$) and our TRGB distance to measure a SFR of $SFR_{FUV}=(2.4\pm0.9)\times10^{-7}\:M_{\odot}\:yr^{-1}$, using Equation 10 from \citet{McQuinn2015}. Adopting \CVnC{}'s present-day stellar mass of $\rm M_*=3.4^{+4.2}_{-2.6}\times10^6\:M_{\odot}$ (Section \ref{sec:lum_mass}), we calculate a specific SFR of log$_{10}(\rm sSFR\:[yr^{-1}])=-13.2^{+0.6}_{-0.4}$; according to the criteria of log$_{10}(\rm sSFR\:[yr^{-1}])<-11$ in \citet{Wetzel2012}, this classifies CVnC as a quenched galaxy. This corroborates the evidence of quenching noted above based on \CVnC{}'s CMD (Section \ref{sec:sfh_match}).

\subsubsection{Gas Content}\label{sec:gas_content}

While the photometric properties of a galaxy can provide important constraints on its recent star formation, its neutral gas content is a crucial tracer of the galaxy's potential for future star formation and can help determine whether a system is truly quenched or simply quiescent (i.e., with the potential of reigniting its star formation).

\CVnC{}'s gas content has been studied in a few H\textsc{i} surveys, the most recent of which include the Systematically Measuring Ultradiffuse Galaxies survey (SMUDGEs; \citealt{Karunakaran2024}) and a survey of nearby dwarf galaxies with the Green Bank Telescope \citep{Nazarova2025}. \CVnC{} exhibits a lack of detectable neutral hydrogen in both surveys. We use upper limits on \CVnC{}'s H\textsc{i} flux from these surveys and our TRGB distance to obtain an upper limit on the H\textsc{i} mass of the galaxy, using relations from \citet{Karunakaran2024}. The more stringent upper limit of the two is given by \citet{Nazarova2025}, which suggests that \CVnC{} has a gas mass of $\rm M_{\text{H}\textsc{i}}<1.5\times10^6\:M_{\odot}$. With \CVnC{}'s stellar mass (Section \ref{sec:lum_mass}), this indicates that \CVnC{} is gas-poor ($\rm M_{\text{H}\textsc{i}}/M_{*}<0.44$). This upper limit on \CVnC{}'s gas content provides further evidence for the system's quenched nature and suggests that \CVnC{}'s lack of recent star formation (Sections \ref{sec:sfh_match} and \ref{sec:sfh_fuv}) is the result of a definitive halt to its star formation, rather than a brief pause. However, we cannot rule out a very low-mass ($<10^6\:M_{\odot}$) gas reservoir without deeper H\textsc{i} follow-up.

\subsection{The Integrated Luminosity and Present-day Stellar Mass of \CVnC{}} \label{sec:lum_mass}

To obtain the integrated luminosity of \CVnC{}, we use the best-fit SFH and distance for the galaxy. Using an MC approach, we first draw a synthetic stellar population using \CVnC{}'s best-fit SFH and a distance sampled from its TRGB distance measurement and associated uncertainties. We then randomly sample from this synthetic population until we are able to replicate the number of stars ($N_*$) in \CVnC{}'s observed CMD (consisting of all of 401 sources which pass our quality cuts and are within $\rm<2R_e$ of the galaxy center). In each iteration, the F606W magnitude of the sampled stars is corrected for extinction, converted to V-band magnitudes using bolometric corrections from \citet{Chen2019}, and summed to obtain the total V-band magnitude, $\rm M_V$. We use the median of these measurements as our reported magnitude, with 16th and 84th percentiles as errors. Using this approach, we obtain a V-band magnitude of $\rm M_V = -11.2^{+0.6}_{-0.5} \:(stat.)^{+0.8}_{-0.4} \:(sys.)$ mag for \CVnC{}. For conciseness, we report the final integrated luminosity of \CVnC{} with the larger of the systematic and statistical uncertainties, $\rm M_V = -11.2^{+0.8}_{-0.5}$ mag.

\begin{figure*}
\begin{subfigure}[t]{0.5\textwidth}
    \includegraphics[width=1.0\linewidth, alt={Two scatter plots showing the structural properties of \CVnC{} in relation to other dwarf galaxies. The first plot shows that the effective radius vs.\ integrated magnitude distributions of satellite and field galaxies overlap significantly but satellite galaxies tend to have higher radii at a given magnitude (i.e., lower surface brightnesses) than field galaxies. \CVnC{} sits on the the upper side of the distribution, and its properties match those of satellite galaxies more than field galaxies. The second plot shows the central surface brightness vs.\ effective radius for a subset of galaxies from the first plot; \CVnC{} sits in the lower-right of the diagram (large effective radius and low surface brightness), in a region mostly uninhabited by other galaxies.}]{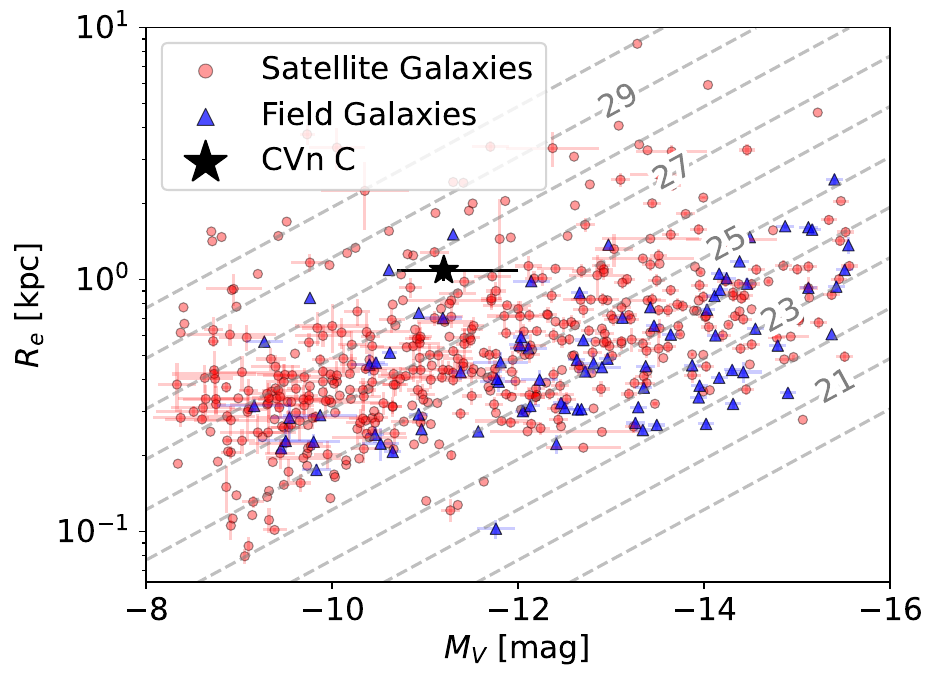}
\end{subfigure}
\hspace{-5pt}
\begin{subfigure}[t]{0.5\textwidth}
    \includegraphics[width=0.94\linewidth]{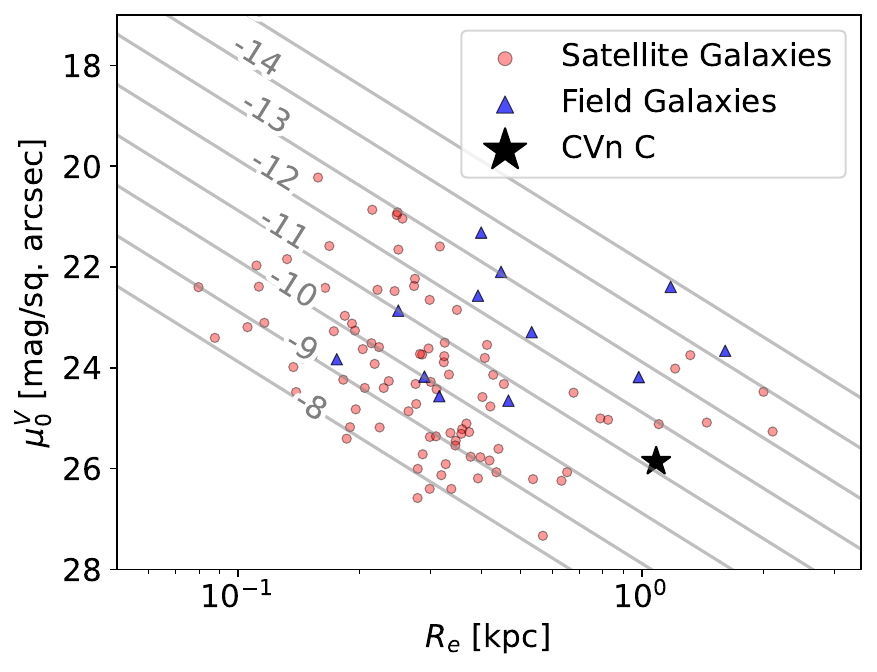}
\end{subfigure}
\caption{Left: Effective Radius vs.\ V-band Magnitude for \CVnC{} (black star) and dwarf galaxies in the relevant mass regime ($\rm M_*=10^{5.5}-10^{8.5}\:M_{\odot}$) from the Local Volume Database \citep{Pace2024}. Satellites are presented with red circles, while field galaxies (with no recorded host in the LVDB) are shown with blue triangles. We show 1$\sigma$ error bars for systems with cataloged uncertainties. Dashed lines show different surface brightness contours in units of mag/sq.\ arcsec. Right: Central V-band Surface Brightness vs.\ Effective Radius for \CVnC{} (black star) and systems from the first plot which have cataloged S\'ersic indices and surface brightnesses. Dashed lines show different absolute magnitude contours in units of mag, assuming the same S\'ersic index as \CVnC{}. \CVnC{} is a low surface brightness galaxy and has properties more consistent with the satellite dwarf galaxy sample than with the field galaxy population. \CVnC{} is also much more extended and has a lower surface brightness than galaxies with similar V-band luminosities in the comparison sample, although this may be partially attributable to detection biases.}
    \label{fig:cvnc_mass_size}
\end{figure*}

The first plot in Figure \ref{fig:cvnc_mass_size} presents the effective radius vs.\ V-band magnitude of \CVnC{} (black star) alongside that of dwarf galaxies in the relevant mass range ($\rm M_*=10^{5.5}-10^{8.5}\:M_{\odot}$) from the Local Volume Database \citep{Pace2024}. \CVnC{} has a low surface brightness and shows properties more consistent with satellites (red circles) rather than field dwarf galaxies (blue triangles).
The second plot in Figure \ref{fig:cvnc_mass_size} displays the central surface brightness vs.\ effective radius of \CVnC{} and galaxies in the first plot which have cataloged S\'ersic indices and surface brightnesses, and which we used to estimate their central surface brightness. \CVnC{} sits on the edge of the galaxy distribution, and exhibits a much more extended radius and lower central surface brightness than galaxies with similar V-band luminosities. However, we note that the dearth of galaxies in this low surface brightness/large radius regime is likely partially attributable to the difficulty in detecting less concentrated galaxies at a fixed surface brightness. 

The stellar mass of \CVnC{} was measured in two different ways. First, the MC approach used above to measure \CVnC{}'s integrated luminosity was similarly used to estimate the galaxy's mass by summing the masses of the $N_*$ stars sampled in each iteration. Using this approach, we obtain a measurement of $\rm M_*=1.2\pm0.4 \:(stat.) \pm0.4\:(sys.)\times 10^6\:M_{\odot}$. Second, we use the best-fit SFH to estimate \CVnC{}'s stellar mass. Our approach integrates the galaxy's SFH over its lifetime to obtain its total stellar mass.
This total stellar mass is then corrected for MATCH's lower IMF normalization, as recommended by \citet{Telford2020}; MATCH integrates the SFH using a Kroupa IMF with no mass cutoffs, i.e., between 0 and $\infty$, leading to a higher stellar mass than would be measured using physically-motivated mass cutoffs (i.e., 0.1-100 $\rm M_{\odot}$). Thus, we apply a correction to measure the stellar mass which would be obtained with a Kroupa IMF integrated over the 0.1-100 $\rm M_{\odot}$ mass range. The measurement is also corrected for the recycling of stellar mass to the ISM, with a recycling fraction of 0.42 based on \CVnC{}'s metallicity ([M/H]=$-$0.6) and Table 2 of \citet{Vincenzo2016}. This yields an estimate of \CVnC{}'s present-day stellar mass of $5.63^{+1.96}_{-3.64}\times10^6\:M_{\odot}$. Our masses from the two methods are reasonably similar but do differ slightly ($<2\sigma$). This may due to imprecision in our fitting of the structural parameters of \CVnC{}, which is a natural consequence of the limited depth of our data. Keeping in mind the uncertainty of this result, we report a final present-day stellar mass of $3.4^{+4.2}_{-2.6}\times10^6\:M_{\odot}$, derived from the average from both methods. Uncertainties reflect the range in mass provided by our two estimates.

\subsection{The Environment Around Canes Venatici C} \label{sec:envir}

\begin{figure*}[t!]
        \centering
        \includegraphics[width=1.0\linewidth, alt={Three-panel scatter plot showing three projections of the galaxies within 1.5 Mpc of \CVnC{} in Supergalactic X, Y, and Z coordinates. The size of the points indicate each galaxy's virial radius, which does not intersect that of \CVnC{} in any of the three projections. NGC 4631 and its satellites sit near the edge of the 1.5 Mpc circle around \CVnC{}.}]{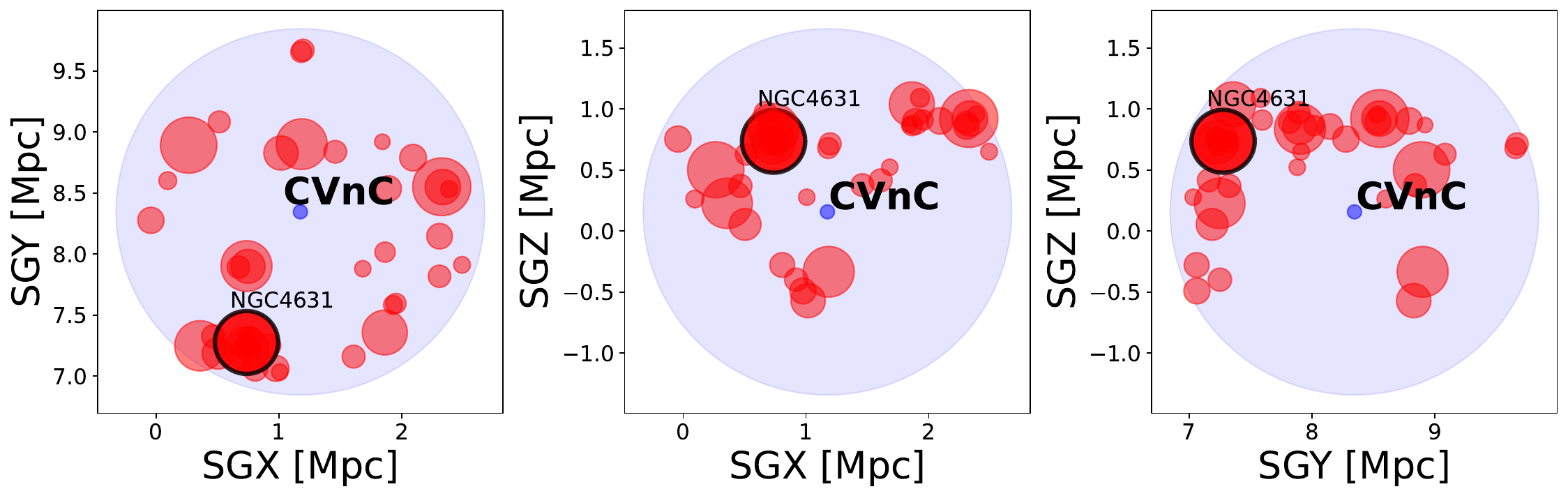}
    \caption{The location of \CVnC{} in Supergalactic coordinates with all identified neighboring galaxies within a physical distance of 1.5 Mpc, indicated by the large light blue circle. The size of the red circles represent the virial radii of each neighbor, while the virial radius of \CVnC{} is shown with a dark blue circle. The location of \CVnC{}’s most influential $L_{\star}$ neighbor (NGC 4631) is annotated and marked with a black outline. \CVnC{} is outside the virial radius of all neighbors, and is located at a 3D separation of $\sim1.3$ Mpc ($>5R_\text{vir}$) from NGC 4631.}
    \label{fig:environ}
\end{figure*}

We probe the environment around \CVnC{} using the TRGB distance of 8.43 Mpc and galaxy catalogs from the Extragalactic Distance Database \citep{Tully2009}. We find 43 galaxies within a 3D separation of 1.5 Mpc, shown in Figure  \ref{fig:environ}. 
This sample was limited to galaxies with secure distance estimates, i.e., using resolved stars (e.g., RR Lyrae, TRGB, HB etc.), as well as systems with less accurate distances based on e.g., group membership, or velocity-based distances calculated from the Numerical Action Method (NAM; \citealp{Shaya2017})\footnote{For all galaxies with Tully-Fisher (TF) or Baryonic Tully-Fisher (BTF) distances, we used their radial velocities to measure NAM distances using the \citet{Kourkchi2020} NAM Distance-Velocity calculator. This was done before using the catalogs to search for \CVnC{}'s neighbours in order to ensure a more complete sample.}.

We assess \CVnC{}'s environment in two ways: we both quantitatively estimate the extent of \CVnC{}'s isolation through its tidal index $\Theta_5$ (explained below) and we qualitatively investigate the possible influence of individual neighboring galaxies and groups.

First, we describe our quantitative measurement of \CVnC{}'s isolation. To estimate each neighbor's influence on \CVnC{}, we measure its tidal force $F$ using the definition from \citet{Karachentsev2013}, i.e., $F\sim M/D^3$. $M$ refers to the mass of the neighbor---assumed to be proportional to its luminosity---and $D$ is the spatial separation from the neighbor. We use stellar masses for $M$, which are based on apparent K-band magnitudes from the UNGC (corrected for Galactic and internal extinction using relations from \citealp{Karachentsev2013}), the galaxies' distances, and an assumed mass-to-light ratio of $\rm M_*/L_k\simeq1$ from \citet{Bell2001}. For an overall estimate of \CVnC{}'s isolation, we measure the tidal index $\Theta_5$---which is based on the five neighbors with the greatest tidal force on \CVnC{} and is calculated as follows, from \citet{Karachentsev2013}:

\begin{equation}
    \Theta_5 =  log(\sum^5_{n=1} M_n/D_n^3)+C
\end{equation}

\noindent where $C=-10.96$ is chosen such that galaxies with $\Theta_5<0$ are likely isolated and those with $\Theta_5>0$ correspond to group members. We measure $\Theta_5=-0.20$ for \CVnC{}, placing it in a relatively isolated environment.

Having quantitatively determined \CVnC{}'s extent of isolation, we turn to individual neighboring galaxies which may have influenced its quenching. Based on our stellar mass estimates, there are 8 massive galaxies ($\rm M_* \geq 10^9 M_{\odot}$) within 1.5 Mpc of \CVnC{}, the closest of which (NGC 4656 and NGC 4062) are $0.7-0.9$ Mpc from \CVnC{}. 
Three of the massive neighbors are $L_{\star}$ galaxies ($\rm M_* > 10^{10} M_{\odot}$\footnote{We employ the lower limit on stellar mass used by \citet{Robotham2013} to select their sample of $L_{\star}$ galaxies.}), all three of which are $>1.1$ Mpc from \CVnC{}. Of the $L_{\star}$ neighbors, the system with the largest tidal force and largest assumed influence on \CVnC{} is the massive ($\rm M_* = 2.7\times 10^{10} M_{\odot}$) galaxy NGC 4631. NGC 4631 and NGC 4656 are also part of the documented NGC 4631 galaxy group \citep{Garcia1993}.

In Figure \ref{fig:environ}, we present the environment within 1.5 Mpc of \CVnC{} in Supergalactic X, Y, and Z coordinates. The size of the red circles show the virial radii of \CVnC{}'s neighboring galaxies, calculated from our stellar mass estimates and the stellar-mass to halo-mass relation from \citet{Behroozi2019}. We find that none of the virial radii of these neighboring galaxies encompass \CVnC{}. Moreover, \CVnC{} is at a separation of $\sim1.3$ Mpc, or $\rm>5R_\text{vir}$, from its most influential $L_{\star}$ galaxy neighbor, NGC 4631. \CVnC{}'s less massive $L_{\star}$ neighbors (NGC 4490 and NGC 4559) are at similarly large separations of $\rm\sim5-6\:R_\text{vir}$. The NGC 4631 galaxy group ($\rm M_{halo}=10^{12.1}\:M_{\odot}$)\footnote{We estimate this group halo mass by summing the halo masses of NGC 4631 and the neighboring galaxies within its virial radius.} likely has the most environmental influence on the system, but \CVnC{}'s large separation from the group puts it in an isolated environment and precludes it from being a conventional satellite which is currently being influenced by the host halo. In Section \ref{sec:discuss}, we discuss mechanisms which may be responsible for \CVnC{}'s quenching, including the possibility that \CVnC{} has had interactions with the NGC~4631 group in the past.

\section{Discussion} \label{sec:discuss}

As we have shown above, \CVnC{} is a relatively isolated ($\Theta_5=-0.20$), low-mass ($\rm M_*=3.4\times10^6\:M_{\odot}$) dwarf galaxy with its most influential $L_{\star}$ neighbor separated by a distance of 1.3 Mpc. The stellar population of \CVnC{} is predominantly old; results from our study of its resolved stellar populations are consistent with little to no star formation in the last 1 Gyr. Independent constraints from the FUV magnitude of \CVnC{} and its H\textsc{i} non-detection from the literature provide a very low specific SFR ($sSFR=10^{-13.2}\:yr^{-1}$) and a stringent upper limit on its gas content ($\rm M_{\text{H}\textsc{i}}<1.5\times 10^6 \:M_{\odot}$, or $\rm M_{\text{H}\textsc{i}}/M_*<0.44$). These pieces of evidence suggest that \CVnC{} is both isolated and quenched. In this section, we discuss possible quenching mechanisms for \CVnC{}.

\subsection{Past Interactions with a Massive Halo} \label{sec:backsplash}

While \CVnC{} is not currently in the halo of the NGC~4631 group, its properties may be explained by past interactions with a massive host. In particular, \CVnC{} may have passed through the NGC~4631 group halo and subsequently left the virial boundary, either through (i) a flyby interaction with the host, (ii) an ejection from the host halo, or (iii) on an extreme backsplash orbit. The first two possibilities would imply that \CVnC{} is not currently bound to the host, while a backsplash orbit would indicate that it is indeed a bound satellite of the halo (albeit with a large apocentric radius). Such dwarf galaxies outside the virial radius of a massive host which are `associated' with the halo, i.e., have passed through its virial volume in the past, are expected to be quite common; \cite{Teyssier2012} use simulations to predict that $13\%$ of all LG field dwarfs have been in the host halo of the MW in the past, and can be found out to distances of 5$R_\text{vir}$.

The hypothesis that \CVnC{} may have had past interactions with a massive host halo is supported by the galaxy's position in size-mass space as well as its low surface brightness.
Dwarf galaxies which have been in the virial volume of a host halo experience environmental processing (e.g., tidal stripping) very similar to conventional satellites, leading to present-day properties consistent with the satellite population \citep{Wetzel2014,Simpson2018}. Our results for \CVnC{} are consistent with this expectation; Figure \ref{fig:cvnc_mass_size} shows that \CVnC{}'s large size for its luminosity is more consistent with the satellite population than the field population of dwarf galaxies. Additionally, \CVnC{} is more extended and has a lower central surface brightness than galaxies with similar V-band luminosities, which may indicate tidal effects \citep[e.g.,][]{Smercina2025}. Thus, \CVnC{}'s present-day properties are consistent with the scenario wherein it experienced satellite-like environmental processing (such as tidal stripping) in NGC 4631's halo in the past.

One may argue that \CVnC{}'s large separation from NGC 4631 precludes it from being an associated galaxy; however, \CVnC{}'s distance from its most influential $L_{\star}$ neighbor can be explained by an extreme backsplash orbit. Smaller backsplash orbits (with apocentric radii $r_{apo}<3R_\text{vir}$) could be natural consequences of anisotropies in the host's mass distribution \citep{Sales2007}. While these smaller orbits are more likely, simulations also predict sizable ($\sim6\%$) instances of extreme backsplash orbits $r_{apo}>4.5R_\text{vir}$ caused by the tidal dissolution of pairs \citep{Sales2007} or groups \citep{Ludlow2009} of subhalos in a massive host halo. In an extreme backsplash scenario, the tidal field of the massive halo can disrupt a satellite group or pair accreted together, redistributing energy among the members such that less massive galaxies are sent on extreme energetic orbits while more massive galaxies are kept on conventional orbits due to higher dynamical friction. Such a scenario was applied to the isolated LG dSph Tucana which, at a separation of $>1.3$ Mpc, may be an extreme backsplash satellite of M31 \citep{Santos-Santos2023}. NGC 4631 is at a similar distance from \CVnC{} ($\sim1.3$ Mpc) and \CVnC{} is less massive than all of NGC 4631's satellites depicted in Figure \ref{fig:environ}, suggesting that one or more of these massive satellites could have been `parent' satellites which were accreted with \CVnC{} and resumed conventional orbits after the group was dissolved by NGC 4631's tidal field.

Another piece of evidence supporting the role of \CVnC{}'s broader environment in its quenching is that its most influential $L_{\star}$ neighbor is the host of the documented NGC 4631 group ($\rm M_{halo}=10^{12.1}\:M_{\odot}$). Group environments have been found to be important in the stripping of gas from satellites \citep{Putman2021}. Large-scale environment is also expected to be important in the regulation of star formation; massive hosts of backsplash systems reside mostly in cluster-sized halos \citep{Bhattacharyya2025}, and dwarf galaxies which form in dense environments form their stars earlier \citep{Gallart2015, Christensen2024}. If \CVnC{} was in a group environment in the past, this may have aided its possible exit from the virial radius of NGC 4631 as well as its early star formation and quenching.

Although our circumstantial evidence supports the idea that CVn~C has had past interactions with NGC~4631, kinematic information would be required to confirm it and distinguish an unbound trajectory from a backsplash orbit. 

\subsection{Other Mechanisms}
\label{sec:other_mech}

Briefly, we discuss other mechanisms which are less likely but nevertheless should be considered in the quenching of \CVnC{}.

Cosmic reionization and stellar feedback may have quenched the lowest-mass galaxies \citep[$\rm M_*\lesssim10^5\:M_{\odot}$, $\rm M_{halo}\lesssim10^9\:M_{\odot}$; e.g., ][]{Wheeler2015,Kim2024}, yet there is evidence that this mass threshold may be closer to $\rm M_*\sim10^4\:M_{\odot}$\citep{McQuinn2023,McQuinn2024c}. However, reionization and feedback is unlikely to have quenched \CVnC{}; \CVnC{}'s stellar mass ($\rm M_{*}=3.4\times 10^6 \:M_{\odot}$) is significantly larger than the expected threshold below which we would expect the star formation in a galaxy to be significantly impacted by this mechanism. 

Another plausible explanation for CVn C’s present-day properties is that we are observing it during an extended period of quiescence driven by outflows of gas via feedback from bursty star formation. This has been employed as a possible formation mechanism for Ultra-diffuse galaxies or UDGs, which are dwarf galaxies with unusually large sizes and low surface brightnesses \citep[e.g.,][]{DiCintio2017}. We note that CVn C’s large effective radius ($\rm R_e>1$ kpc) and low central surface brightness ($\rm\mu_0^V\sim26\: mag/arcsec^2$) is consistent with the broadly defined population of UDGs \citep[e.g.,][]{Yagi2016,DiCintio2017}, although CVn C is somewhat less extended than the standard \citet{VanDokkum2015} definition of a UDG ($>1.5$ kpc). As a UDG, CVn C’s lack of star formation could be explained by these “breathing-mode” outflow cycles. However, \citet{DiCintio2017} predict that field UDGs formed via cyclic feedback-driven outflows should have significant HI gas masses ($\sim10^{7-9}\: M_{\odot}$), and gas availability is crucial to this formation mechanism. CVn C’s stringent upper limit on HI mass ($\rm\lesssim10^6\:M_{\odot}$) thus likely precludes this mechanism from explaining the system’s properties.

Another possibility is that \CVnC{} has been quenched via cosmic web-stripping. Impulsive ram-pressure from the cosmic web can strip the gas from low-mass halos \citep{Benitez-Llambay2013}. This has been posed as a possible explanation for the signs of ram-pressure stripping seen in the isolated dwarf WLM \citep{Yang2022,Cohen2025}, while stripping via cosmic web filaments could explain the anisotropic quenching signal seen in galaxy clusters at large distances ($\sim2.5R_\text{vir}$) from the halo \citep{Stephenson2025}. Cosmic web-stripping is predicted to be especially efficient at lower masses ($\rm M_*\lesssim10^7\:M_{\odot}$) due to shallower potential wells \citep{Benavides2025}, making it a plausible quenching mechanism for \CVnC{}.

While both past interactions and cosmic web-stripping scenario are plausible for \CVnC{}, we lack the information to conclusively determine whether it is one (or either) of these possibilities. Quenching timescales of backsplash and cosmic web-stripped galaxies as populations are distinct but unlikely to provide a clean test to disentangle the mechanisms for an individual galaxy \citep{Benavides2025}, especially since these are highly dependent on orbits and dynamical friction timescales for satellites and backsplash systems \citep{Hirschmann2014}. The ratios of stellar and gas mass to dynamical mass could help distinguish between the two scenarios; the quenching of systems in a host halo acts through tidal stripping and leads to higher $\rm M_*/M_{dyn}$ and $\rm M_{\text{H}\textsc{i}}/M_{dyn}$ and scatter in relations between stellar, halo, and gas mass \citep{Wetzel2014,Bhattacharyya2025}, while ram-pressure stripping via the cosmic web is not expected to show similar scatter \citep{Benavides2025}. However, obtaining a measurement of the dynamical mass of \CVnC{} is difficult; \CVnC{}'s lack of H\textsc{i} content prohibits a measurement of the velocity profile of the galaxy via its gas, while its distance prevents us from measuring its stellar velocity dispersion. Even so, constraints on \CVnC{}'s dark matter content may be possible in the future with deeper H\textsc{i} follow-up data (assuming \CVnC{} has gas below the current non-detection limit) or through the measurement of \CVnC{}'s Globular Cluster Mass Function (GCLF).

\section{Conclusions}
\label{sec:conclusion}

In this paper, we present new HST imaging of the resolved stars in the dwarf galaxy \CVnC{}. Using the new data in tandem with archival constraints and existing galaxy catalogs, we show that \CVnC{} appears to be an isolated, quenched dwarf galaxy in the Local Volume. Our main takeaways are noted below.

\begin{enumerate}
    \item Using HST data of \CVnC{}, we measure its structural parameters and find that it has a large size ($R_e\sim1$ kpc) for its luminosity ($\rm M_V = -11.2$ mag), which is more comparable to the properties of satellites than field dwarf galaxies. We also find that \CVnC{} is more extended and has a lower central surface brightness than nearby dwarf galaxies with similar V-band luminosities, which may indicate the presence of tidal stripping.
    \item We construct a CMD of the galaxy's resolved stellar population. We see a clear RGB feature when compared to a representative background region, with a small number of possible AGB stars. We use isochrones and CMD-fitting to deduce that the \CVnC{} has a predominantly old stellar population, with most stars aged between 1-10 Gyr and no evidence of young ($<100$ Myr) stars. However, we are unable to rule out a small amount of late-time (100 Myr to 1 Gyr) star formation.
    \item We use our HST-based CMD to find a TRGB distance to \CVnC{} of $8.43^{+0.47}_{-0.32}$ Mpc, which places the galaxy in an isolated environment. \CVnC{}'s most influential $L_{\star}$ galaxy neighbor is NGC 4631, at a separation of $\sim1.3$ Mpc, or $>5R_\text{vir}$.
    \item We independently quantify \CVnC{}'s recent star formation and gas mass using our TRGB distance to \CVnC{}, in tandem with archival constraints on the system's FUV flux and H\textsc{i} content. \CVnC{} has a low specific SFR (log$_{10}(sSFR\:[yr^{-1}])=-13.2$) and has no detected gas ($\rm M_{\text{H}\textsc{i}}/M_{*}<0.44$), supporting its classification as a quenched system.
    \item We investigate possible mechanisms which may have quenched \CVnC{}, and find circumstantial support for past interactions with a host halo, e.g., an extreme backsplash orbit by the tidal dissolution of a subhalo group. However, other mechanisms (e.g., quenching via cosmic web stirpping) cannot be ruled out with our current data.
\end{enumerate}

Further constraints on \CVnC{} may be possible with deeper H\textsc{i} follow-up to detect a low-mass gas reservoir, if present. Kinematic information from spectroscopic follow-up, as well as an estimate of the system's dynamical mass through \CVnC{}'s GCLF, could also help distinguish between different quenching scenarios presented here. Although difficult at the distance of \CVnC{}, a deeper investigation into the intriguing properties of quenched, isolated dwarf galaxies can bring critical insights. Studying such objects can put constraints on the masses at which different quenching mechanisms operate, disentangle complex environmental effects, and help us understand the evolutionary pathways of dwarf galaxies.

\begin{acknowledgments}

The contributions of each author are as follows. T.N.H. led the overall analysis and writing of the paper. K.B.W.M. led the HST program to obtain the data presented here, and aided in the analysis and writing of the paper. K.B.W.M., Y.-Y.M., and D.S. developed the search algorithm that detected \CVnC{}. Y.-Y.M., R.E.C., and E.T. aided in the measurement of the properties of \CVnC{} and its environment. E.T. also aided in the writing of the paper. J.A.B., A.E.D., and M.J.B.N. provided software and analysis tools to measure the distance to \CVnC{} and derive its photometry and star formation history. A.S. provided guidance in the interpretation of \CVnC{}’s properties.

We would like to express our gratitude to the anonymous referee for their insightful comments, which strengthened our manuscript. Support for this work was provided by NASA through grant no.\ HST-GO-17481, awarded by the Space Telescope Science Institute, which is operated by the Association of Universities for Research in Astronomy, Incorporated, under NASA contract NAS5-26555. This work was also supported by a Cottrell Scholar Award from the Research Corporation for Science Advancement (Award No. CS-CSA-2023-110). T.N.H. is partially supported by the NSF Graduate Research Fellowship, under award no. 2233066. This work has made use of the Local Volume Database\footnote{\url{https://github.com/apace7/local_volume_database }} \citep{Pace2024}. This research has also made use of the  the Astrophysics Data System, funded by NASA under Cooperative Agreement 80NSSC21M00561, and the arXiv preprint server.

\end{acknowledgments}

%

\vspace{5mm}
\facilities{HST (ACS and WFC3)}

\software{\texttt{matplotlib} \citep{Hunter2007}, \texttt{numpy} \citep{Walt2011}, \texttt{astropy} \citep{2013A&A...558A..33A,2018AJ....156..123A}, \texttt{drizzlepac} \citep{Hack2013,Avila2015}, \texttt{APLpy} \citep{aplpy2012,aplpy2019}, \texttt{DOLPHOT} \citep{Dolphin2000,Dolphin2016}, \texttt{MATCH} \citep{Dolphin2002,Dolphin2012,Dolphin2013}
          }

\bibliography{main}
\bibliographystyle{aasjournalv7}

\end{document}